\documentclass[aps,pre,twocolumn,showpacs,superscriptaddress]{revtex4-1}

\usepackage{amsmath}
\usepackage{amssymb}
\usepackage{bbold} 
\usepackage{mathtools}
\usepackage{graphicx}
\usepackage{hyperref}
\usepackage[arrow, matrix, curve]{xy}
\usepackage{bm}
\usepackage{bbm} 
\usepackage{yhmath}
\usepackage{color}
\usepackage{url}
\newcommand\diff{\mathrm{d}}

\usepackage[usenames,dvipsnames]{xcolor}
\hypersetup{colorlinks=true, linkcolor=BrickRed, urlcolor=blue!50!black, citecolor=blue!50!black}

\usepackage[normalem]{ulem}

\makeatletter
\newcommand\hide@visible[1]{%
  \bgroup\fboxsep=.3ex\colorbox{Gray}{begin hide}%
  #1\colorbox{Gray}{end hide}\egroup%
}
\newcommand\hide@hidden[1]{%
  \bgroup\fboxsep=.3ex\colorbox{Gray}{hidden text}%
}
\newcommand\hide@invisible[1]{}
\newcommand\makevisible{\let\hide\hide@visible}
\newcommand\makehidden{\let\hide\hide@hidden}
\newcommand\makeinvisible{\let\hide\hide@invisible}
\makeatother
\makehidden

\usepackage[capitalise,nameinlink]{cleveref}

\crefname{section}{Sec.}{Secs.} 
\usepackage{graphicx}
\usepackage[caption=false]{subfig} 
\usepackage{color}
\usepackage{verbatim} 
\usepackage{natbib}

\begin{document}

\title{Intermediate scattering function of a gravitactic circle swimmer}
\author{Regina Rusch}
\affiliation{Institut f\"ur Theoretische Physik, Technikerstra{\ss}e 21-A, Universit\"at Innsbruck, A-6020 Innsbruck, Austria}
\author{Oleksandr Chepizhko}
\affiliation{ Lakeside Labs GmbH, Lakeside B04 b, A-9020 Klagenfurt, Austria}
\author{Thomas Franosch}
\affiliation{Institut f\"ur Theoretische Physik, Technikerstra{\ss}e 21-A, Universit\"at Innsbruck, A-6020 Innsbruck, Austria}

\date{\today}

\begin{abstract}
	We analyze gravitaxis of a Brownian circle swimmer by deriving and characterizing analytically the experimentally measurable intermediate scattering function (ISF). To solve the associated Fokker-Planck equation we use a spectral-theory approach and find formal expressions in terms of eigenfunctions and eigenvalues of the overdamped-noisy-driven-pendulum problem. We further perform a Taylor series of the ISF in the wavevector to read off the cumulants up to the fourth order. We focus on the skewness and kurtosis analyzed for four observation directions in the 2D-plane. Validating our findings involves conducting Langevin-dynamics simulations and interpreting the results using a harmonic approximation. The skewness and kurtosis are amplified as the orienting torque approaches the intrinsic angular drift of the circle swimmer from above, highlighting deviations from Gaussian behavior. Transforming the ISF to the comoving frame, again a measurable quantity, reveals gravitactic effects and diverse behaviors spanning from diffusive motion at low wavenumbers to circular motion at intermediate and directed motion at higher wavenumbers. 
\end{abstract}

\maketitle

\section{Introduction}

Microswimmers, microscopic objects capable of moving in liquid environments,  have attracted considerable interest because of their important role in nature, such as in fertilization processes~\cite{Nosrati_2015, schwarz2020sperm, debnath2020enhanced}, or in diverse biomedical applications, for example, micro motors~\cite{Baraban_2012, Jahir_2013, Koumakis_2013, Liu_2020}, imaging~\cite{pane2019imaging, hosseini2020high}, microsurgery~\cite{vyskocil2020cancer}, targeted drug delivery~\cite{Yan_2022, Mahon_2012,Tottori_2012}, and many more~\cite{bunea2020recent}. Of interest is their nontrivial dynamics, stemming from their intrinsic state of being out of equilibrium because of the ongoing conversion of energy into directed motion. 
Microswimmers can be found in a variety of systems, both biological, such as bacteria, sperm cells, and microorganisms, and artificially synthesized, including Janus particles, magnetic microswimmers, and active colloids~\cite{kurzthaler2024characterization,Zhiyu_2021,Bechinger_2016,Elgeti_2015}.  
In particular, circle swimmers move in curved trajectories via interactions with their physical shape, propulsion mechanism, symmetry, or external interactions ~\cite{Liebchen_2017, Kuemmel_2013, lowen2016chirality, BorgetenHagen_2014}.

Taxis, in all its forms, embodies the ability of organisms and particles to actively respond to external stimuli, for instance, this capability could enable microorganisms to avoid sedimentation or adjust to the intensity of the light they are exposed to. Their guided motions can be influenced by external factors, such as chemical gradients, light intensity, magnetic fields, fluid flows, or gravity - referred to as gravitaxis. Negative gravitaxis, which is the response to move oppositely to the gravitational field has been observed for organisms, such as \emph{Euglena gracilis}~\cite{Lebert_1999}, \emph{Paramecium}~\cite{Hemmersbach_2001} or asymmetric self-propelled colloidal particles~\cite{BorgetenHagen_2014}. Theoretical modeling and experiments on these colloidal particles demonstrated their alignment because of gravitational forces, forming the foundation for our current model. In a previous study~\cite{Chepizhko_2022}, two of the present authors used the L-shaped gravitactic circle swimmer from Ref.~\cite{BorgetenHagen_2014} and mapped it to an overdamped noisy driven pendulum. This allowed computing analytically the variance and diffusivity, which revealed a resonant diffusivity when the orienting torque approaches the intrinsic angular drift of the circle swimmer~\cite{Chepizhko_2022}. While the variance provides insight into transport properties ranging from directed and circular motion to diffusive behavior, extracting further information by computing higher moments becomes increasingly more tedious. For the case of vanishing external force, higher-order moments and correlation functions have been derived analytically using a Laplace transform-based method applied to the Fokker-Planck equation~\cite{Kurzthaler_2017,pattanayak2024impact}. 

The ISF encodes all moments in terms of derivatives with respect to the wave vector, providing full spatio-temporal information and has been computed analytically for an anisotropic active Brownian particle~\cite{Kurzthaler_2016,Kurzthaler_2018_janus} and  Brownian circle swimmer~\cite{Kurzthaler_2017}. Furthermore, the ISF is directly accessible via experiments, for example,  dynamic light scattering~\cite{VanMegen_1991}, single particle tracking~\cite{Kurzthaler_2018_janus} or differential dynamic microscopy (DDM)~\cite{Cerbino_2008,Wilson_2011}.

In this work, we shall analytically derive the ISF for the model of a circle swimmer subjected to gravity, as described in Ref.~\cite{Chepizhko_2022} across various length and time scales. We analyze the gravitactic effects by comparing different strengths of the orienting torque relative to the intrinsic angular drift. To validate our findings, we will conduct Langevin-dynamics simulations and rationalize the results in terms of a harmonic approximation, focusing on various regimes near the bifurcation point.
We solve the corresponding Fokker-Planck equation in Fourier space by using an operator representation and a spectral-theory approach. The analytical expressions involve the eigenvalues and eigenfunctions of this operator, which can be computed numerically. Furthermore, we derive analytically the skewness and kurtosis by time-dependent perturbation theory. This method has previously demonstrated to be applicable in various contexts, such as the anisotropic active Brownian particle~\cite{Kurzthaler_2016} or the anisotropic Brownian circle swimmer~\cite{Kurzthaler_2017}. In our current work, we extend this analytical framework to encompass microswimmers navigating an external field, with a specific focus on the phenomenon of gravitaxis~\cite{BorgetenHagen_2014,Chepizhko_2022}.

\section{Model} \label{sec: model}
\begin{figure}[tb] 
	\centering
	\includegraphics[width=0.5\textwidth]{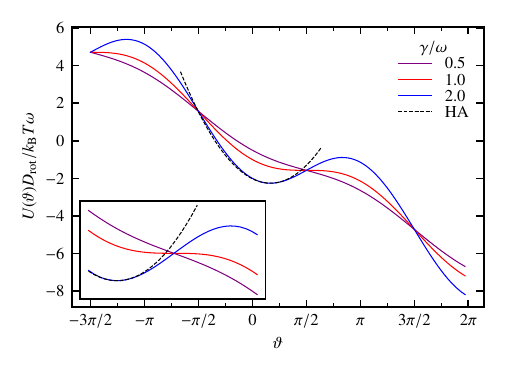}
	\caption{Tilted washboard potential for the angle $\vartheta$ and different torques $\gamma $ above, exactly at, and below the classical bifurcation. The harmonic approximation (HA) for the case of $\gamma /\omega = 2$ is shown in dashed lines. The inset is a zoom close to the inflection point.}
	\label{fig:washboard_potential}
\end{figure}

An active L-shaped particle moving in the presence of an external gravitational field causing a torque has been introduced in Ref.~\cite{BorgetenHagen_2014}. It has been demonstrated that the asymmetry in its shape alone was sufficient to induce gravitactic behavior. We use a simplified model where noise terms and additional drift components associated with (anisotropic) translational diffusion are omitted. In principle, one can include these terms as outlined in the Supplemental Material of Ref.~\cite{Chepizhko_2022}, however they do not qualitatively affect the circular motion of interest for discussing gravitaxis. The particle's position in space  
\begin{align} \label{eq:positioneom}
	\dot{\textbf{r}}(t)= v \textbf{u} (t)=v 
	\begin{pmatrix}
		\cos \vartheta(t) \\ \sin \vartheta(t)
	   \end{pmatrix} ,
 \end{align}
evolves in a plane with constant speed $v$ along its orientation parametrized by an angle $\vartheta$. 
The angular dynamics are governed by the equation of motion 
\begin{align} \label{eq:thetaeom}
	\dot{\vartheta }(t) = \omega - \gamma \sin \vartheta (t) + \zeta (t).
\end{align}
Here $\omega >0$ denotes the constant angular drift because of the shape asymmetry. The gravitational force acting on the particle translates to an orienting torque, $\gamma  \propto g, \gamma > 0$, via the translational-rotational coupling, see the Supplemental Material of Ref.~\cite{Chepizhko_2022}.
Furthermore, $\zeta (t)$ corresponds to a centered Gaussian white noise with delta-correlated variance $\langle \zeta (t) \zeta (t^\prime) \rangle = 2 D_\text{rot} \delta (t-t^\prime)$, where $D_\text{rot}$ is the rotational diffusion coefficient. 
Strictly speaking, $\gamma$  has dimensions of a drift term, nevertheless we refer to it as a torque since it tends to align the L-shaped particle because of gravity. As a peculiarity, gravity causes the angle $\vartheta$ to align in the $x$-direction perpendicularly to the gravitational field. The L-shaped particle is oriented along $\vartheta$ and an additional Hall angle, similar to the Hall effect~\cite{BorgetenHagen_2014,Chepizhko_2022}.

There is an instructive mechanical analogy to the overdamped noisy driven pendulum~\cite{stratonovich_1967_topics, gitterman_2008_noisy}. The only difference is that the angle in the pendulum problem is measured from the negative $y$-axis. In this analogy, $\omega$ represents a driving motor trying to increase the angle at a constant rate. In contrast, gravity parametrized by $\gamma $ aligns the pendulum to a downward position.
The deterministic part of the angular motion in \cref{eq:thetaeom} derives from an effective potential

\begin{align} \label{eq: washboard_potential}
	\frac{D_\text{rot}}{k_\text{B}T} U(\vartheta)= - \omega \vartheta - \gamma \cos \vartheta ,
\end{align} 
 see \cref{fig:washboard_potential}. As all angles are $2 \pi$-periodic, periodic boundary conditions are applied. 
For $\gamma > \omega $ the potential displays a local minimum located at $\vartheta_* = \arcsin {(\omega /\gamma ) }\in [0, \pi/2]$. Similarly, the potential exhibits a maximum at $\pi - \vartheta_*$, yielding a barrier height $\Delta U = U(\pi - \vartheta_*)- U(\vartheta ^*) $. If $\gamma =\omega $, the potential displays an inflection point. In the absence of noise, this point corresponds to a saddle-node bifurcation. In the pendulum analogy, above this point ($ \gamma > \omega $), there is a stable fixed point at $\vartheta_*$ where the angle is locked. In contrast, below the bifurcation point ($\gamma < \omega$), no fixed points exist and the motor drives the pendulum to complete full rotations. Correspondingly, we refer to $\gamma > \omega$ as the locked state and to $\gamma < \omega$ as the running state~\cite{strogatz2018nonlinear}.
The introduction of noise smears this transition. 

Another special case that is analytically tractable is the Brownian circle swimmer, i.e., without the orienting torque, which has been already fully characterized~\cite{Kurzthaler_2017,gitterman_2008_noisy}.

\section{Derivation of the intermediate scattering function}
Besides the equations of motion for the individual trajectories from \cref{sec: model}, dynamical properties are equivalently encoded in the propagator $\mathbb{P} \coloneq \mathbb{P}(\mathbf{r}, \vartheta ,t | \vartheta_0)$, which is defined as the conditional probability distribution of the particle displaced by $\mathbf{r}$ at lag time $t$ with orientation $\vartheta $, given it started with orientation $\vartheta_0$.
By standard means~\cite{risken1996fokker}, the associated Fokker-Planck equation can be derived 
\begin{align} 
	\partial_t {\mathbb{P}} &= - \partial_\vartheta [(\omega - \gamma \sin \vartheta ){\mathbb{P}}] + D_\text{rot} \partial^2_\vartheta \mathbb{P} - v \textbf{u} \cdot \partial_\mathbf{r}   {\mathbb{P}} .
\end{align}
The initial condition is specified by $\mathbb{P}(\mathbf{r},\vartheta, t=0|\vartheta_0)= \delta (\vartheta - \vartheta_0) \delta (\mathbf{r})$. Here, all angles are considered to be $2\pi$-periodic. 

It is favorable to solve directly for the spatial Fourier transform of the propagator $\tilde{\mathbb{P}} \coloneq  \tilde{\mathbb{P}} (\mathbf{k}, \vartheta ,t | \vartheta_0) $
\begin{align}
	\tilde{\mathbb{P}} =  \int_{\mathbb{R}^2} \! \diff \,\textbf{r} \exp (- \textsf{i} \textbf{k} \cdot \textbf{r}) \mathbb{P}(\mathbf{r}, \vartheta ,t | \vartheta_0).
\end{align}
The Fourier-transformed Fokker-Planck equation then reads
\begin{align} \label{eq:fokker-planck}
	\nonumber \partial_t \tilde{\mathbb{P}} &= - \partial_\vartheta [(\omega - \gamma \sin \vartheta )\tilde{\mathbb{P}}] + D_\text{rot} \partial^2_\vartheta \tilde{\mathbb{P}} - \textsf{i} v \textbf{k} \cdot \textbf{u}\tilde{\mathbb{P}}  \\
	&\eqcolon (\mathcal{L} + \delta \mathcal{L}_\mathbf{k}  )\tilde{\mathbb{P}}, 
\end{align}
where the operator $\mathcal{L}$ encodes the angle-dependent motion containing the drifts $\omega $ and $\gamma$, and the rotational diffusion. The $\mathbf{k}$-dependent operator $\delta \mathcal{L}_\mathbf{k}$ encodes the deterministic active motion for a given wave vector $\mathbf{k}$. The initial condition then follows $\tilde{\mathbb{P}}(\mathbf{k},\vartheta, t=0|\vartheta_0)= \delta (\vartheta - \vartheta_0)$. Hence, $\tilde{\mathbb{P}}$ encodes the full information on the translational and angular motion.

In the subsequent analysis we focus on the translational motion. The ISF $F(\textbf{k},t)\coloneq \langle \exp(-\textsf{i}\textbf{k}\cdot\Delta\textbf{r}(t)) \rangle $ is the characteristic function of the stochastic displacements $\Delta \textbf{r}(t) \coloneq \textbf{r}(t)-\textbf{r}(0)$ encoding the spatio-temporal information for the process resolved at time scale $t$ and length scale $2\pi/|\mathbf{k}|$. Furthermore, by taking derivatives with respect to the wave vector the low-order moments can be obtained. 

The ISF is derived by averaging over the initial angle with the steady-state probability density $p^{\text{st}}(\vartheta_0) $ and marginalizing over the angle $\vartheta$
\begin{align} \label{eq:ISF1}
	F(\textbf{k},t) = \int_0^{2\pi}\!\diff \vartheta \int_0^{2\pi}\!\diff \vartheta_0\, \tilde{\mathbb{P}} (\mathbf{k}, \vartheta ,t | \vartheta_0) p^{\text{st}}(\vartheta_0).
 \end{align}

The formal solution of \cref{eq:fokker-planck} is $\tilde{\mathbb{P}}  =  e^{(\mathcal{L} + \delta \mathcal{L}_\mathbf{k} ) t }\delta (\vartheta - \vartheta_0)$ and to obtain explicit expressions a spectral-theory approach is applied on the operators. We shall also employ a time-dependent perturbation theory for small wave vectors $\mathbf{k}$ to generate the low-order moments. The following derivation is already outlined in Ref.~\cite{Chepizhko_2022} (see also~\cite{Lapolla_2020, Kurzthaler_2017}) and for completeness, we summarize the most important steps.

We use the Dirac notation and solve the eigenvalue problem of the system $\mathcal{L}$ as well as of the full system $\mathcal{L}+\delta \mathcal{L}_\mathbf{k}$ and write the operators in terms of the abstract Hilbert space representation. Therefore, we introduce the scalar product 
\begin{align} \label{eq:scalar_product}
	\langle f|g \rangle =  \int_0^{2\pi} \!\diff\vartheta\, f(\vartheta )^*g(\vartheta ) = \int_0^{2\pi}  \!\diff\vartheta\,\langle f|\vartheta \rangle  \langle \vartheta |g \rangle,
\end{align} 
where we use the generalized angular basis $\{ | \vartheta \rangle \}$ which is orthogonal in the following sense $\langle \vartheta | \vartheta_0 \rangle = \delta (\vartheta - \vartheta_0)$ and (over-) complete $\int_0^{2\pi} \diff  \vartheta |\vartheta \rangle\langle \vartheta| =  \mathbb{1}$. We rely on the isomorphism between the periodic square-integrable functions $f(\vartheta) , g(\vartheta) \in L^2[0,2\pi]$ and states $|f\rangle $,$|g\rangle $ in the Hilbert space  $\mathcal{H}$. 
We are interested in the eigenfunctions $\{ | \alpha \rangle : \alpha \in \mathbb{Z}\} $ of the unperturbed operator $\mathcal{L}$ for the standard orthonormal basis in $\mathcal{H}$, with the real-space representation $ \langle \vartheta| \alpha \rangle = \exp ( \textsf{i} \alpha \vartheta) /\sqrt{2\pi}$. The unperturbed operator $\mathcal{L}$ can be constructed in this basis by its matrix elements
\begin{align} \label{eq:matrix_representation_L}
	\mathcal{L}_{\beta \alpha} &= \langle \beta| \mathcal{L}\alpha \rangle \coloneq \int_0^{2\pi} \!\frac{\diff \vartheta}{2 \pi}\,\exp{(-\textsf{i} \beta \vartheta)} \mathcal{L} \exp{(\textsf{i} \alpha \vartheta )}  \\
	&=\delta_{\beta \alpha}  (-D_{\text{rot}}\beta^2-\textsf{i}\beta \omega ) + \frac{\gamma	}{2}  (\beta \delta_{\beta,\alpha+1}- \beta \delta_{\beta,\alpha-1}) .\nonumber 
\end{align}
The form of the resulting matrix is tridiagonal if the external torque is non-zero, and in general the matrix becomes  non-Hermitian. 
Since the operator has the symmetry, $\mathcal{L}_{(-\beta)(-\alpha)} = \mathcal{L}^*_{\beta \alpha} $,  $ \sum\nolimits_\alpha |\alpha \rangle \langle -\alpha| r_n \rangle ^*$ is a right-eigenvector to eigenvalue $\lambda^*$ if  $|r \rangle = \sum\nolimits_\alpha |\alpha \rangle \langle \alpha | r \rangle$ is a right-eigenvector to eigenvalue $\lambda $.  A similar relation holds for the left-eigenvectors and eigenvalues, see \cref{sec:matrix_elements} for the derivations.
The eigenvalue $\lambda_0=0$ represents the stationary state and the corresponding eigenstates are called $| r_0 \rangle$ and $\langle l_0 |$. Using \cref{eq:matrix_representation_L} one can directly observe that the left eigenstate to eigenvalue zero is $\langle l_0 | = \langle 0 | $ with the real space representation $ l_0(\vartheta) = \langle l_0 | \vartheta \rangle  = 1$. The right eigenstate to eigenvalue zero is the stationary state with real space representation $r_0(\vartheta) = \langle \vartheta | r_0 \rangle = p^{\text{st}} (\vartheta)  $, because of the normalization of the probability density. 
Thus, we can define $\textit{right}$ and $\textit{left}$ eigenstates $\mathcal{L} |r_n \rangle = -\lambda_n |r_n \rangle $, $\mathcal{L}^\dagger |l_n \rangle = -\lambda_n^* |l_n \rangle $, respectively and label them by $n \in \mathbb{Z}$.  Here, $\mathcal{L}^\dagger $ denotes the adjoint of $\mathcal{L}$ with respect to the scalar product \cref{eq:scalar_product}.  Thus, $\lambda_{-n}= \lambda_n^*$ and $\langle \alpha| r_{-n} \rangle = \langle -\alpha| r_n \rangle ^*$
The eigenstates are chosen to be orthonormal $ \langle l_m | r_n \rangle = \delta_{m,n}$ and assumed to fulfill the completeness relation $\sum_n |r_n \rangle \langle l_n |=  \mathbb{1}$.

The formal solution of the \cref{eq:fokker-planck} can be rewritten in the eigenbasis
\begin{align}
	\tilde{\mathbb{P}}  =   \langle \vartheta |e^{(\mathcal{L} + \delta \mathcal{L}_\mathbf{k} ) t } \vartheta_0\rangle ,
\end{align}
and for the ISF follows
\begin{align}\label{eq:ISF2}
	F(\textbf{k},t) &= \int_0^{2\pi}\! \diff \vartheta \int_0^{2\pi} \! \diff  \vartheta_0\,\langle l_0 | \vartheta \rangle \langle \vartheta |e^{(\mathcal{L} + \delta \mathcal{L}_\mathbf{k} ) t } \vartheta_0\rangle   \langle \vartheta_0 | r_0 \rangle \nonumber \\
	&=  \langle l_0  |e^{(\mathcal{L} + \delta \mathcal{L}_\mathbf{k} ) t }  r_0 \rangle .
\end{align} 
Here, we used the abstract notation for the stationary state and inserted a  $\langle l_0 | \vartheta \rangle =1$ and used the completeness relations for $\vartheta$ and $\vartheta_0$.

Building upon the model presented in Ref.~\cite{Chepizhko_2022}, summarized above, we derive an explicit expression for the ISF in terms of the eigenfunction and eigenvalues of the full operator. Therefore, we apply now the same formalism to the full operator $\mathcal{L}+ \delta \mathcal{L}_\mathbf{k}$ and find equivalently to \cref{eq:matrix_representation_L}  the missing matrix elements 
\begin{align} \label{eq:matrix_representation_deltaL}
	&( \delta \mathcal{L}_\mathbf{k})_{\beta \alpha} = \langle \beta |  \delta \mathcal{L}_\mathbf{k} \alpha \rangle  \nonumber \\
	&=- \frac{ \textsf{i} k_x v }{2} (\delta_{\beta ,\alpha +1}+\delta_{\beta ,\alpha-1}) 
	- \frac{k_y v }{2}(\delta_{\beta ,\alpha+1}-\delta_{\beta ,\alpha-1}), 
\end{align}
 which results again in a tridiagonal matrix with left and right eigenstates and eigenvalues
\begin{align} \label{eq:eigenvalues_full_operator}
	(\mathcal{L} +\delta \mathcal{L}_\mathbf{k}  )|r_{n \mathbf{k}} \rangle = -\lambda_{n \mathbf{k}} |r_{n \mathbf{k}}  \rangle , \\
	(\mathcal{L}^\dagger +\delta \mathcal{L}^\dagger_\mathbf{k}  )|l_{n \mathbf{k}} \rangle = -\lambda^*_{n \mathbf{k}} |l_{n \mathbf{k}}  \rangle .
\end{align}
 Consequently, the time evolution operator can be expressed as the spectral sum of the eigenvalues and eigenstates of the full operator
\begin{align}
	e^{(\mathcal{L} +\delta \mathcal{L}_\mathbf{k}  )t} = \sum_{n \in \mathbb{Z}} e^{-\lambda_{n \mathbf{k}}t}  |r_{n \mathbf{k}}  \rangle  \langle l_{n \mathbf{k}}  | ,
\end{align}
which leads to the final representation of the ISF 
\begin{align} \label{eq:intermediate_scattering_function}
	F(\textbf{k},t) & = \sum_{n \in \mathbb{Z}}  e^{-\lambda_{n \mathbf{k}}t}  \langle l_0 |r_{n \mathbf{k}}  \rangle  \langle l_{n \mathbf{k}}  | r_0 \rangle . 
\end{align}
This allows us to compute the ISF as a sum of decaying exponentials with the eigenvectors and eigenvalues of the matrix representations of both, $\mathcal{L}$ and $\mathcal{L}+\delta \mathcal{L}_\mathbf{k}$. Since the eigenvalues can become complex, the ISF may display oscillations. This property is in striking contrast to equilibrium systems where the dynamics is always \emph{completely monotone} \cite{feller1991introduction}.

\section{Harmonic Approximation} \label{sec:harmonic_approximation}
The effective potential, \cref{eq: washboard_potential}, shows a local minimum $\vartheta_* $ in the locked phase, $ \gamma > \omega $, where a harmonic approximation (HA) can be applied, see \cref{fig:washboard_potential}. This approximation holds in the regime of small fluctuations $D_\mathrm{rot} \ll \omega $ given fixed potential barriers with $\gamma > \omega $, not too close to the bifurcation. In Ref.~\cite{Chepizhko_2022}, the harmonic approximation is derived by approximating the Fokker-Planck equation, here, in contrast, we show the derivation in the Langevin picture. From there we also derive the ISF in the harmonic approximation. We shall also discuss the limits of the HA. 

Considering a small perturbation $\delta \vartheta (t)$ around the fixed angle $\vartheta_* : \vartheta(t) = \vartheta_*  + \delta \vartheta (t)$ the linearized equation of motion of \cref{eq:thetaeom} for this perturbation reads
\begin{align} \label{eq:langevin_HA}
	\frac{\diff }{\diff t} \delta \vartheta (t) = -\frac{1}{\tau} \delta \vartheta (t)+\zeta (t),
\end{align}
with the characteristic time scale $1/\tau = \sqrt{\gamma^2 - \omega^2}$. In this approximation $\delta \vartheta(t)$ becomes a Gaussian random process with zero mean. Its dynamics are characterized completely by its autocorrelation function 
\begin{align}
	\langle \delta \vartheta(t)  \delta \vartheta(0)  \rangle = D_\text{rot} e^{-|t|/\tau} .
\end{align}
The dynamics in space simplify to
\begin{align}
		\frac{\diff }{\diff t}  \mathbf{r} (t) = v  \begin{pmatrix}
			\cos \vartheta_* \\ \sin \vartheta_*
		   \end{pmatrix} + v \delta \vartheta_*(t)  \begin{pmatrix}
			-\sin \vartheta_* \\ \cos \vartheta_*
		   \end{pmatrix} ,
\end{align}
and from there the displacement in space is computed 
\begin{align} 
	 \Delta \mathbf{r} (t)  &=  \mathbf{r} (t) -  \mathbf{r} (0) \nonumber \\
	 &=  v t\begin{pmatrix}
		\cos \vartheta_* \\ \sin \vartheta_* 
	   \end{pmatrix} + v \begin{pmatrix}
		-\sin \vartheta_*  \\ \cos \vartheta_*
	\end{pmatrix}\int_0^t\!\diff s\, \delta \vartheta_*(s).
\end{align}
Therefore, also the displacement becomes a Gaussian processes completely characterized by its mean and covariance. Then
\begin{align}\label{eq:ha_mean_displacement}
	\mathbf{n} \cdot  \langle \Delta \mathbf{r} (t)  \rangle  &=    v t \mathbf{n}\cdot  \begin{pmatrix}
		\cos \vartheta _* \\ \sin \vartheta_* 
	   \end{pmatrix} ,
\end{align}
 corresponds to the motion in the locked phase ignoring fluctuations. The observation direction is set by $\mathbf{n} \coloneq \mathbf{k}/k$.
\begin{figure}[bt] 
	\centering
	\includegraphics[width=0.5\textwidth]{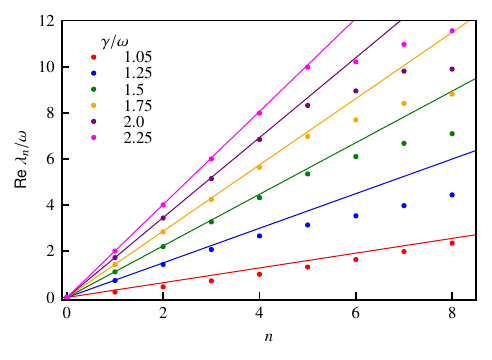}
	\caption{Real part of the lowest eigenvalues $\lambda_n$ for $n \geq 0$ of the unperturbed operator $\mathcal{L}$ for $D_\mathrm{rot}/\omega =0.005$ for increasing values of $\gamma$ above the bifurcation. Points correspond to numerical values, and lines display the harmonic approximation $\lambda^\text{HA}_n$.}
	\label{fig:eigenvalues}
\end{figure}
Similarly, we derive the variance by averaging over the squared, centered mean 
\begin{align}
	&\textsf{Var} [\mathbf{n} \cdot   \Delta \mathbf{r} (t)  ]  =  \langle [\mathbf{n} \cdot \Delta \mathbf{r} (t) -\mathbf{n} \cdot  \langle \Delta \mathbf{r} (t)\rangle ]^2\rangle  \nonumber \\ 
	&= \left[ v  \mathbf{n} \cdot 
	\begin{pmatrix}
		-\sin \vartheta _* \\ 
		\cos \vartheta_* 
	\end{pmatrix} 
	\right]^2 \int_0^t \!\diff t_1 \int_0^t \! \diff t_2\, \langle \delta \vartheta(t_1)  \delta \vartheta(t_2) \rangle.
\end{align}
We obtain again the same expression for the harmonic approximation of the variance as in Ref.~\cite{Chepizhko_2022} 
\begin{align} \label{eq:HA_variance}
	&\textsf{Var} [\mathbf{n} \cdot   \Delta \mathbf{r} (t)  ]  = 2 D_\mathbf{n} ( t - \tau + \tau e^{-t/\tau} ),
\end{align}
with
\begin{align}
	D_\mathbf{n}= (v\tau)^2 D_\text{rot} (n_x \sin \vartheta_* - n_y \cos \vartheta_* )^2.
\end{align}
As no higher cumulants exist in harmonic approximation, the expression for the ISF can be derived
\begin{align} \label{eq:HA_ISF}
	F(\mathbf{k},t) =  \exp (-\textsf{i} k\mathbf{n} \cdot  \langle \Delta \mathbf{r} (t)  \rangle - \frac{k^2}{2} \textsf{Var} [\mathbf{n}\cdot  \Delta \mathbf{r} (t)   ]  ),
\end{align}
indicating that oscillations in the ISF stem from the mean displacement of the particle, while the strength of the exponential decay is determined by the variance. Both observables will be discussed in the following sections \cref{sec:exact_low_order,sec:directions}.

The eigenvalues for harmonic motion are known~\cite{risken1996fokker}: they are real, discrete and evenly spaced $\lambda^\mathrm{HA}_n =n/\tau$ with  $n\in \mathbb{N}_0$. In \cref{fig:eigenvalues} we illustrate how well the real part of the lowest eigenvalues, \textsf{Re}~$\lambda_n$, derived numerically from the unperturbed operator $\mathcal{L}$, \cref{eq:matrix_representation_L}, coincides with the eigenvalues of the harmonic approximation.
As for $n \neq 0$ the eigenvalues show up in pairs with the same real part, we plotted the real part once, i.e. the eigenvalues with positive imaginary part $n \geq 0$.

\section{Mean motion and directions of observation}
\label{sec:directions}
\begin{figure}[bt] 
	\centering
	\includegraphics[width=0.5\textwidth]{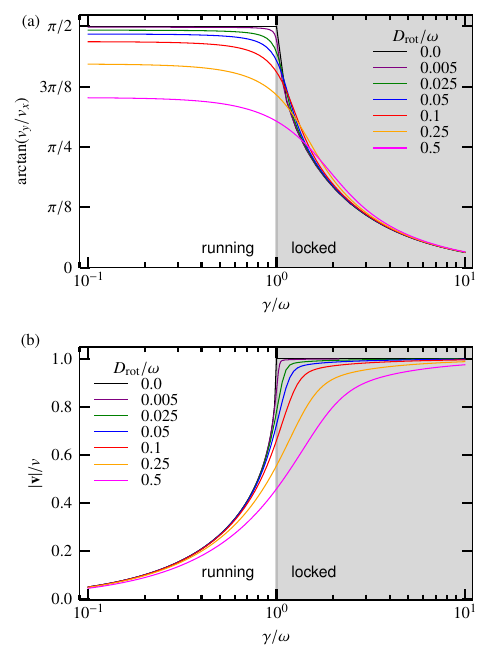}
	\caption{
		(a) The angle $\arctan (v_y/v_x)$ of the direction of motion and (b) the absolute value of the mean velocity $|\mathbf{v}|$  for various rotational diffusion constants $D_\text{rot}$. The black line corresponds to the deterministic motion $D_\text{rot}/\omega=0$.  }
	\label{fig:mean_displacement_D_0_025}
\end{figure}
The gravitactic circle swimmer model comprises four parameters: The radius of the circular motion $v/\omega$ denotes the unit of length, while the unit of time is represented by $1/\omega$. The dimensionless torque is defined as $\gamma / \omega$ and the relative significance of fluctuations is quantified as $D_\mathrm{rot} /\omega$. Characteristic times are the time needed for completing one circle $\tau_\omega=2 \pi/ \omega$ and the rotational diffusion time $\tau_\mathrm{rot}=1/D_\mathrm{rot}$, in which the particle loses its orientation. We are interested in the case $\tau_\omega \ll \tau_\text{rot}$, in which the particle's orientation becomes randomized long after it completes one or more circles. 

It is instructive to discuss the mean motion for the case of vanishing fluctuations first. The results for this case are known and can also be found in~\cite{risken1996fokker}. For the case $\gamma > \omega $ a fixed point exists and the solution follows the harmonic approximation as described in \cref{eq:ha_mean_displacement}. When $\gamma \leq \omega $ we solve the classical equation of motion
\begin{align}
	\dot{\vartheta} = \omega - \gamma \sin \vartheta,
\end{align} 
and compute the period by averaging over time steps $\mathrm{d}t$
\begin{align}
	T=\int_0^{2 \pi}\! \frac{\mathrm{d} \vartheta}{\omega-\gamma \sin \vartheta}=\frac{2 \pi}{\sqrt{\omega^2-\gamma^2}},
\end{align}	  
which diverges as $\gamma \uparrow \omega$. The average drift is determined by averaging over the velocity 
\begin{align}\label{eq:deterministic_mean_velocity}
	\langle \mathbf{v} \rangle &=\frac{1}{T} \int_0^T \!\mathrm{d} t \,v 
		\begin{pmatrix}
		\cos \vartheta(t)\\
		\sin \vartheta(t)
		\end{pmatrix}
		=\frac{v}{T} \int_0^{2 \pi} \!\frac{\mathrm{d} \vartheta}{\omega-\gamma \sin \vartheta} \begin{pmatrix}
			\cos \vartheta\\
			\sin \vartheta
			\end{pmatrix}\nonumber \\
			&=v \begin{pmatrix}
					0\\
					\omega / \gamma-\sqrt{\left.(\omega / \gamma)^2-1\right)}
				\end{pmatrix} .
\end{align}
The particle drifts on average along the $y$-direction, opposite to gravity.

We will now discuss the case with fluctuations and the mean motion along certain directions of observation denoted by $\mathbf{n}$. Natural choices are the direction parallel to gravity, represented by $\mathbf{n}_y = (0,1)^T$, as well as its perpendicular counterpart $\mathbf{n}_x = (1,0)^T$. Additionally, we look at directions aligned with the mean motion and perpendicular to it. From Ref.~\cite{Chepizhko_2022} the mean motion along $\mathbf{n}=\mathbf{k}/k$ of the problem is known
\begin{align}
	&\mathbf{n} \cdot\frac{\diff }{\diff t}  \langle \Delta \mathbf{r}(t)   \rangle = \mathbf{n} \cdot \mathbf{v}= \frac{\textsf{i}}{k} \langle l_0 |  \delta\mathcal{L}_{\mathbf{k}}   r_0 \rangle.  \label{eq:meanvelocity}
\end{align} 
The mean velocity $\mathbf{v}= (v_x,v_y)^T=v(\langle\cos \vartheta \rangle, \langle \sin \vartheta \rangle)^T$ can be inferred by choosing the observation directions as $\mathbf{n}_x$ and $\mathbf{n}_y$, respectively. The mean of these trigonometric functions can also be calculated directly from the knowledge of the stationary distribution $p^\text{st}(\vartheta )$ of the angular motion, compare Ref.~\cite{risken1996fokker}. We introduce the directions parallel and perpendicular to the mean velocity as 
\begin{align}
	\mathbf{n}_\parallel  &= \frac{1}{|\mathbf{v}|}
	\begin{pmatrix}
		v_x \\
		v_y
	   \end{pmatrix},  \qquad 
	\mathbf{n}_\perp  = \frac{1}{|\mathbf{v}|}
	\begin{pmatrix}
		-v_y \\
		v_x
	   \end{pmatrix}.
\end{align}

The direction of the particle motion is given by the angle $\arctan (v_y/v_x)$. For low rotational diffusion $D_\text{rot}$ and below the classical bifurcation, $\gamma \leq \omega $, the particle predominantly moves along the $y$-axis against gravity, corresponding to an angle of $\pi/2$, see \cref{fig:mean_displacement_D_0_025} (a). This behavior was computed in \cref{eq:deterministic_mean_velocity} for the deterministic case. As the rotational diffusion increases, this alignment angle decreases.
In the locked phase, as $\gamma $ increases, the angle increasingly aligns with the $\mathbf{n}_x$ direction since gravity forces the angle in this orientation. The harmonic approximation, represented by $D_\text{rot}/\omega=0$ and defined only in the locked state, aligns more closely with the numerical values as $\gamma $ increases and $D_\text{rot}$ decreases, which is consistent with theoretical expectations.

In the locked phase, the absolute value of the mean velocity $|\mathbf{v}|$ approaches the maximal velocity $v$ for large $\gamma \gg \omega$, see \cref{fig:mean_displacement_D_0_025}$\,$(b). This convergence occurs faster as the rotational diffusion $D_\text{rot} $ decreases.  The better the angle is locked, the faster the velocity becomes.
In contrast, for small $\gamma \ll \omega$, the velocity converges to zero as in this regime there is no preferred direction.

\section{Intermediate scattering function}

We determine numerically the ISF,~\cref{eq:intermediate_scattering_function}, for various values of the wave vector $\mathbf{k} $, covering a range of length scales in terms of the radius of the circular motion.  We also conduct Langevin-dynamics simulations and compare our findings. In principle, this should give identical results, however both methods contain different sources of errors. The simulation errors are of statistical nature, the
numerical error arise from the truncation of the matrix that has to be diagonalized. 

\begin{figure}[tbh] 
	\centering
	\includegraphics[width=0.5\textwidth]{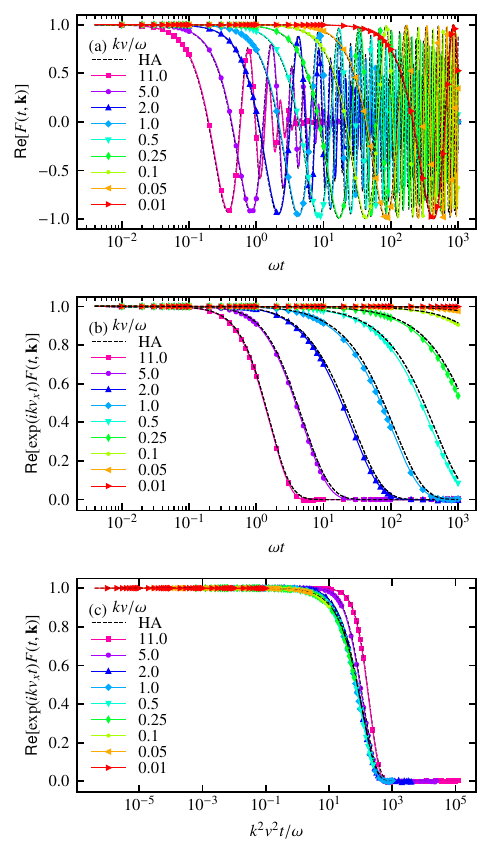}
	\caption{Real part of the ISF for different values of the wave vectors in $x$ direction $\mathbf{k} = k \mathbf{n}_x$ for a diffusion coefficient  $D_\mathrm{rot}/\omega=0.025$ and $\gamma / \omega =1.5$. Full lines correspond to the spectral theory and symbols to Langevin-simulation results. The dashed lines correspond to the harmonic approximation (HA). (a) Real part of the ISF. (b) Same quantity in the comoving frame. (c) Same upon rescaling time  by $k^2v^2/\omega$. }
	\label{fig:ISF_HA_D_0.025_g_1_5}
\end{figure}

For the case of a high gravitational torque $\gamma/\omega =1.5$ and a low diffusion coefficient $D_\mathrm{rot}/\omega=0.025$ the harmonic approximation for the ISF,~\cref{eq:HA_ISF}, holds and is in good agreement with the Langevin-dynamics simulations and numerical results, compare \cref{fig:ISF_HA_D_0.025_g_1_5}.
The direction of observation is chosen to be perpendicular to the gravitational force, i.e., the wave vector takes the form $\mathbf{k}  = k \mathbf{n}_x$. In \cref{fig:ISF_HA_D_0.025_g_1_5} the real part of the ISF is dominated by weakly damped oscillations originating from the drift motion. To remove this effect of the drift, we transform to a frame comoving with the mean velocity, $\exp({i k v_x t})F(\mathbf{k}, t)$. This can also be achieved experimentally using the DDM technique, by shifting the images according to the known mean velocity. The theoretical background is given in \cref{sec:DDM}.
Then the corrected ISF decays monotonically with an increasing decay time as the wavenumber decreases. For small $k \ll v/\omega $ the motion becomes diffusive and the comoving ISF approaches an exponential decay,  $\exp({i k v_x t})F(\mathbf{k}, t) \to \exp(-Dk^2t)$.

The imaginary part of the ISF is not shown here but is as well dominated by oscillations, which again can be eliminated by the transform to the comoving frame. Without the oscillations, the imaginary part is about 2\% of the real part and thus negligible. 

\begin{figure*}[htb] 
	\centering
	\includegraphics[width=\textwidth]{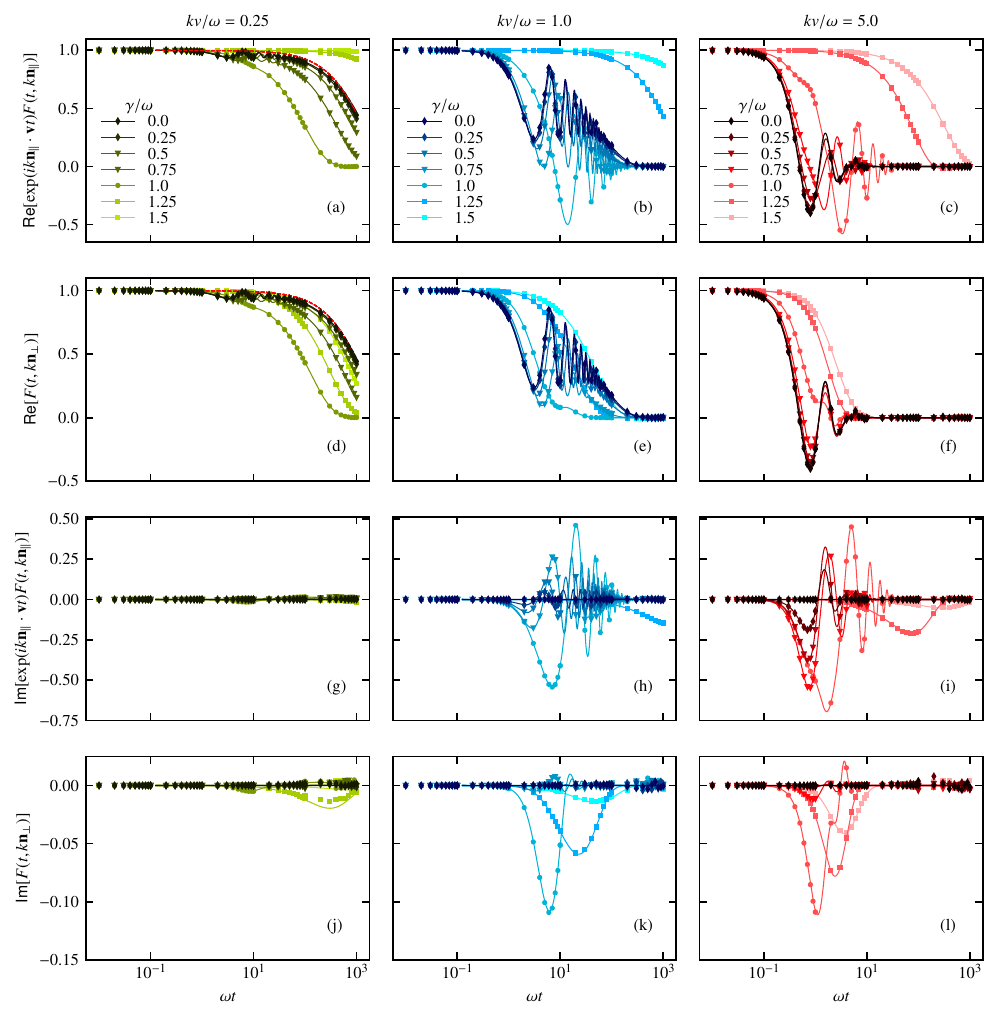}
	\caption{The real and imaginary parts of the ISF in the comoving frame for different orienting torques $\gamma $ and for various values and two directions of the wave vector $\mathbf{k}$. The rotational diffusion coefficient $D_\text{rot}=0.025 \omega$ is chosen. For three different lengths $k=0.25, 1.0,5.0$ there is a plot for two different directions, parallel $\mathbf{n}_\parallel$ ((a)-(c) and (g)-(i)) and perpendicular $\mathbf{n}_\perp$ ((d)-(f) and (j)-(l)) to the direction of the mean velocity. Full lines correspond to the spectral theory and symbols to the Langevin-simulation results. The red dotted line in (a) and (d) correspond to the effective diffusion of a free circle swimmer $\exp(-D_0k^2t)$. Each legend in the first row is valid for the whole column.}
	\label{fig:ISF_comoving_D_0.025}
\end{figure*}

In the following, we analyze the ISF in the comoving frame for the direction of observation parallel and perpendicular to the mean motion. In the perpendicular direction, the ISF in the laboratory and comoving frame are identical since $\mathbf{v} \cdot  \mathbf{n}_\perp =0$. We focus on the case of small fluctuations, $D_\mathrm{rot} \ll \omega$.

Several examples of the real and imaginary parts of the ISF computed using the spectral theory and Langevin simulations are shown in \cref{fig:ISF_comoving_D_0.025} and match each other perfectly. We show sets of the ISF for three wavenumbers $k$, in parallel and perpendicular direction. 
Each panel of the figures displays the ISF arranged in ascending order of the orienting torque $\gamma $ and shows several values below, at, and above the classical bifurcation,~$\gamma =\omega$. 

The darkest line corresponds to the case $\gamma = 0$, no external driving, illustrating the behavior of a Brownian circle swimmer, as already discussed in detail in Ref.~\cite{Kurzthaler_2017}, where the ISF is derived analytically in terms of generalizations of the Mathieu functions. Since without torque the system is isotropic, the ISF is independent of the direction of observation, in particular, the imaginary part evaluates to zero, see \cref{fig:ISF_comoving_D_0.025}. For smaller wavenumbers, the ISF approaches an exponential decay, $\exp(-D_0 k^2t)$, with the effective diffusion coefficient of a free circle swimmer, $D_0= v^2D_\mathrm{rot} /[2(D_\mathrm{rot}^2+\omega^2)]$, see~\cref{fig:ISF_comoving_D_0.025}~(a) and (d). 

Below the classical bifurcation the behavior is similar to the Brownian circle swimmer, although with a mean drift. For large wavenumbers, $k \gg v/\omega$, length scales smaller than the circular motion are probed and the dynamics appear persistent. The oscillations around zero in the ISF originate from this persistent motion.
For intermediate wavenumbers, $k \simeq v/\omega$, the circular motion is resolved which results in oscillations around a finite plateau at time scales $t=\tau_ \omega $.  At times $t  \gtrsim \tau_\text{rot}$, these oscillations start to fade out as rotational diffusion takes over. The oscillations diminish and persist for shorter periods for even smaller wavenumbers.
The oscillations have a bigger amplitude in the parallel direction compared to the perpendicular one.
Approaching the bifurcation the oscillation frequency becomes small as is expected from the motion without fluctuations. 
The imaginary part of the ISF shows oscillations below the bifurcation and those oscillations reach the maximum amplitude at the bifurcation. At small wavenumbers, the imaginary part is negligibly small compared to the real part and increases for larger values of the wavenumber. In the perpendicular direction, the imaginary part has significantly lower amplitudes than the parallel direction.

Above the bifurcation, the real part of the ISF decays monotonically, as the particle's orientation is essentially locked. For small wavenumbers, an exponential decay remains with a diffusion coefficient determined by the variance. The diffusivity is the largest right above the bifurcation and decreases with increasing $\gamma $. 

In the perpendicular direction, the exponential decay occurs at a much faster rate compared to the parallel direction. Consequently, the diffusion coefficient is higher in the perpendicular direction and increases in proximity to the bifurcation point.

\section{Exact Low-Order moments} \label{sec:exact_low_order}

\begin{figure*}[tb] 
	\centering
	\includegraphics[width=\textwidth]{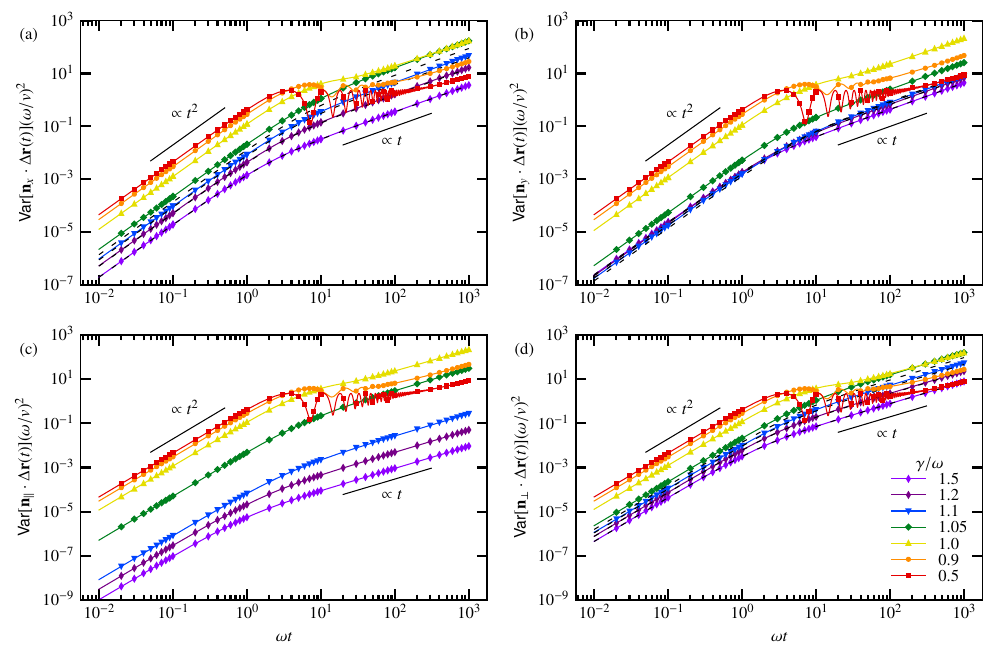}
	\caption{Variance for different values of the orienting torque $\gamma$ and for four directions $\mathbf{n}$ for a rotational diffusion coefficient $D_\text{rot}/\omega = 0.005$. Full lines correspond to the spectral theory and symbols to the Langevin-simulation results. The dashed black lines represent the harmonic approximation (HA) for $\gamma > \omega $. The black solid lines are auxiliary lines to serve as a guide to the eye.  }
	\label{fig:6}
\end{figure*}
The first moments can be derived from the characteristic function of the random displacements, the intermediate scattering function. Our goal is to compute the time-dependent skewness and kurtosis and analyze them for different orienting torques $\gamma$ and directions of observation. 
For completeness, we show the derivation of the mean motion and variance, which have been derived in Ref.~\cite{Chepizhko_2022, risken1996fokker}. 

The skewness is derived similarly, but the cumulant expansion for small wavenumbers $\mathbf{k}=k \mathbf{n}$ has to be extended to higher orders
\begin{align}\label{eq:cumulant_expansion_3order}
	 \ln F(\mathbf{k}, t) =& \sum_{j=1}^{\infty} \frac{(-\textsf{i} k)^j \kappa_j[ \mathbf{n} \cdot \Delta \mathbf{r}(t) ]}{j!}  =  
	 -\textsf{i}  k \mathbf{n} \cdot \langle \Delta \mathbf{r}(t) \rangle \nonumber \\ &-  \frac{ k^2 }{2} \langle [  \mathbf{n} \cdot \Delta\mathbf{r}(t)  - \langle  \mathbf{n} \cdot \Delta \mathbf{r}(t) \rangle ]^2\rangle  \nonumber \\ 
	 & 	+ \frac{ \textsf{i}  k^3}{3!} \langle [  \mathbf{n} \cdot \Delta \mathbf{r}(t) -\langle  \mathbf{n} \cdot \Delta \mathbf{r}(t) \rangle ]^3\rangle  \nonumber\\
	 & 	+ \frac{  k^4}{4!}  \big[ \langle [  \mathbf{n} \cdot \Delta \mathbf{r}(t) -\langle  \mathbf{n} \cdot \Delta \mathbf{r}(t) \rangle ]^4\rangle  \nonumber\\
	 & -3 \langle [  \mathbf{n} \cdot \Delta\mathbf{r}(t)  - \langle  \mathbf{n} \cdot \Delta \mathbf{r}(t) \rangle ]^2\rangle^2 \big] +\ldots . 
\end{align}
Here, $\kappa_j[ \mathbf{n} \cdot \Delta \mathbf{r}(t)]$ represents the $j$th cumulant of the random variable $ \mathbf{n} \cdot \Delta \mathbf{r}(t)$. The goal is to compare the cumulant expansion to the expansion of the ISF and read off the cumulants. Therefore, we expand the ISF, \cref{eq:ISF2}, by iteratively substituting the time evolution operator with the Dyson representation~\cite{sakurai2011modern}
\begin{align} \label{eq:dyson_representation}
	&e^{(\mathcal{L}+\delta \mathcal{L}_{\mathbf{k}} ) t } = e^{\mathcal{L} t} + \int_0^t \!\diff s \, e^{\mathcal{L} (t-s)} \delta \mathcal{L}_{\mathbf{k}} e^{(\mathcal{L} +\delta \mathcal{L}_\mathbf{k} ) s} ,
\end{align} 
up to the fourth order. We refer to the \cref{sec:time_evolution}, where the time evolution operator up to the fourth order is shown. The expansion of the ISF is derived by sandwiching the time evolution operator between the zero eigenvalue states $\langle l_0 | $, $|r_0 \rangle$, see \cref{eq:ISF2}, and then simplifying the expression, equivalently as described in detail in Ref.~\cite{Chepizhko_2022}. Taking the logarithm of the expression yields the cumulant expansion of the ISF, given that the operator $ \delta \mathcal{L}_{\mathbf{k}} $ is of first order in $k$ 
\begin{widetext}
\begin{align} \label{eq:cumulant_expansion_3order_ISF}
	\ln(F(\mathbf{k},t)) =& 	t \langle l_0 |   \delta \mathcal{L}_{\mathbf{k}}   r_0\rangle
	 + 
	\sum_{n \neq 0} \frac{e^{-\lambda_n t} + \lambda_n t -1}{\lambda_n^2}  
	\langle l_0 |\delta \mathcal{L}_{\mathbf{k}}  r_n \rangle \langle l_n |  \delta\mathcal{L}_{\mathbf{k}}   r_0 \rangle  \nonumber \\
	&+ 
	\sum_{n \neq 0}   \frac{\lambda_n  t+e^{-\lambda_n  t} (\lambda_n  t+2)-2}{\lambda_n ^3} 
	\langle l_0 |\delta \mathcal{L}_{\mathbf{k}}  r_n \rangle \langle l_n |  \delta\mathcal{L}_{\mathbf{k}}   r_0 \rangle  (\langle l_n | \delta \mathcal{L}_{\mathbf{k}}  r_n \rangle-\langle l_0 | \delta \mathcal{L}_{\mathbf{k}}  r_0 \rangle 	)   \nonumber \\
	&+
	\sum_{n \neq 0} \sum_{m \neq 0,m \neq n } \left(\frac{e^{-\lambda_n t}+\lambda_n  t-1}{\lambda_n ^2 (\lambda_m -\lambda_n )}  + \frac{e^{-\lambda_n t}+\lambda_n  t-1}{\lambda_m ^2 (\lambda_n -\lambda_m )}\right) \langle l_0 | \delta \mathcal{L}_{\mathbf{k}}   r_n \rangle \langle l_n | \delta \mathcal{L}_{\mathbf{k}}    r_m \rangle \langle l_m | \delta \mathcal{L}_{\mathbf{k}}  r_0 \rangle 	+O(|\mathbf{k}|^4) .
\end{align}
\end{widetext}
Here, all sums run over all integers excluding zero and pairs $m=n$. The fourth-order term is rather lengthy and the calculations are deferred to  \cref{sec:higher_moments}. The first three cumulants   can be read off by comparison to the expansion in \cref{eq:cumulant_expansion_3order}, and we obtain the same expressions for the mean velocity, \cref{eq:meanvelocity}, and variance as in Ref.~\cite{Chepizhko_2022}

\begin{align} \label{eq:variance}
	&\textsf{Var} [\mathbf{n} \cdot \Delta \mathbf{r}(t)  ] = \kappa_2[\mathbf{n} \cdot \Delta \mathbf{r}(t)]=\langle [ \mathbf{n} \cdot \Delta \mathbf{r}(t) -\langle \mathbf{n} \cdot \Delta  \mathbf{r}(t) \rangle ]^2\rangle  \nonumber \\
	&= - \frac{2}{k^2} \sum_{n \neq 0} \frac{e^{-\lambda_n t} + \lambda_n t -1}{\lambda_n^2} 
	\langle l_0 |\delta \mathcal{L}_{\mathbf{k}}  r_n \rangle \langle l_n |  \delta\mathcal{L}_{\mathbf{k}}   r_0 \rangle . 
\end{align}
There are pairs of complex conjugate eigenvalues  $\lambda_n =  \lambda_{-n}^*$, for $n \neq 0$ and it holds that $\langle l_m |  \delta  \mathcal{L}_{\mathbf{k}}  r_n \rangle^* = - \langle l_{-m} |  \delta  \mathcal{L}_{\mathbf{k}}  r_{-n} \rangle $, see \cref{sec:matrix_elements}. Thus,  summing over these pairs in \cref{eq:cumulant_expansion_3order_ISF,eq:variance} results in just the real parts of the terms with even order in $k$. Similarly, to the odd orders in $k$ only the imaginary parts contribute. Comparing to \cref{eq:cumulant_expansion_3order}, we find that because of this symmetry all cumulants are real.

In this work, we extend  the analysis on the total variance as discussed in Ref.~\cite{Chepizhko_2022}. Our investigation includes four directions of observation: the $x$ and $y$ axes, as well as the directions parallel and perpendicular to the mean velocity, see \cref{fig:6}. We note that the total variance can be computed by summing the $x$ and $y$ directions or the parallel and perpendicular directions. Along these directions, we explore various values of the orienting torque $\gamma $, spanning ranges above, exactly at, and below the classical bifurcation.

For small times we find persistent growth that is quadratic in time $\propto t^2$, followed by oscillations starting at times $t \gtrsim \tau_\omega $ for values $\gamma $ below the bifurcation, similar to a free circle swimmer~\cite{Kurzthaler_2017,Kuemmel_2013}.  The oscillations are fading out for long times as the angular motion is randomized, and the particle starts to show diffusive behavior for $t \gtrsim \tau_\text{rot} $, indicated by a linear increase in time. The curves for the case below the classical bifurcation are similar in all four directions, indicating that in the running phase the diffusion becomes isotropic.

At and above the bifurcation we observe diffusive behavior and no oscillations occur as the angle $\vartheta$ is locked by the strong gravitactic torque. Close to the classical bifurcation, $\gamma = \omega$, the diffusivity, i.e., the slope in the long-time regime, is enhanced, as recently discussed in Ref.~\cite{Chepizhko_2022}.

For the case of a strong orienting torque $\gamma >\omega$,  the prefactor of diffusive growth $\propto t$  is much bigger for the perpendicular direction than for the parallel direction, i.e., the particle diffuses mainly in the direction perpendicular to the mean motion. 
In the harmonic approximation this becomes exact as the diffusion coefficient, $D_\textbf{n}=0$, evaluates to zero for the parallel direction, compare \cref{eq:HA_variance}. The approximation coincides with the curves, the better, the stronger the orienting torque. For the case of $\gamma =1.05 \omega$, the harmonic approximation does not match the analytic curve. The reason has been identified in~\cite{Chepizhko_2022} as the particle hops over the barrier before relaxing in the minimum. 

In conclusion, we find that the particle in the locked phase diffuses primarily in perpendicular direction, while in this direction the mean velocity is zero. The closer the particle's mean velocity is to the maximum $v$, the lower is its diffusivity, which can be seen in the parallel direction. In the periodic running phase, the diffusivity becomes isotropic and the mean velocity increases close to the bifurcation. 

\begin{figure*}[tb] 
	\centering
	\includegraphics[width=\textwidth]{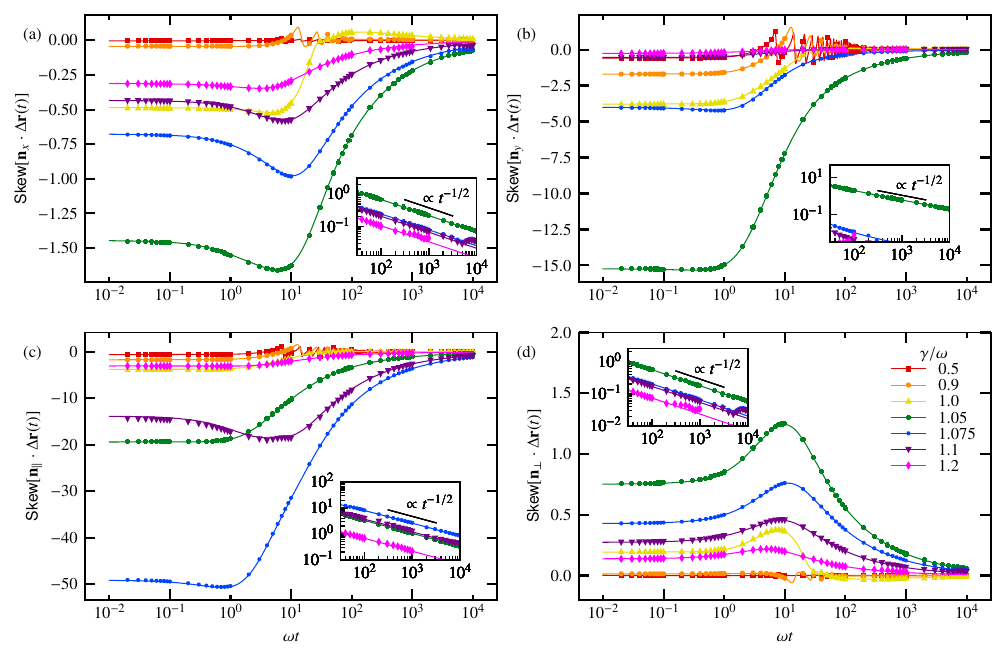}
	\caption{Skewness for different values of the orienting torque $\gamma $ and four directions $ \mathbf{n}$ for a rotational diffusion coefficient $D_\text{rot}/\omega = 0.005$. Full lines correspond to the spectral theory and symbols to the Langevin-simulation results. Inset: The absolute value of the Skewness $|\textsf{Skew}[ \mathbf{n} \cdot \Delta  \mathbf{r}(t) ]|$ on double logarithmic scale to demonstrate the algebraic decay. The black solid lines are auxiliary lines to serve as a guide to the eye. }
	\label{fig:Skewness_D_0.005}
\end{figure*}

\begin{figure*}[tb] 
	\centering
	\includegraphics[width=\textwidth]{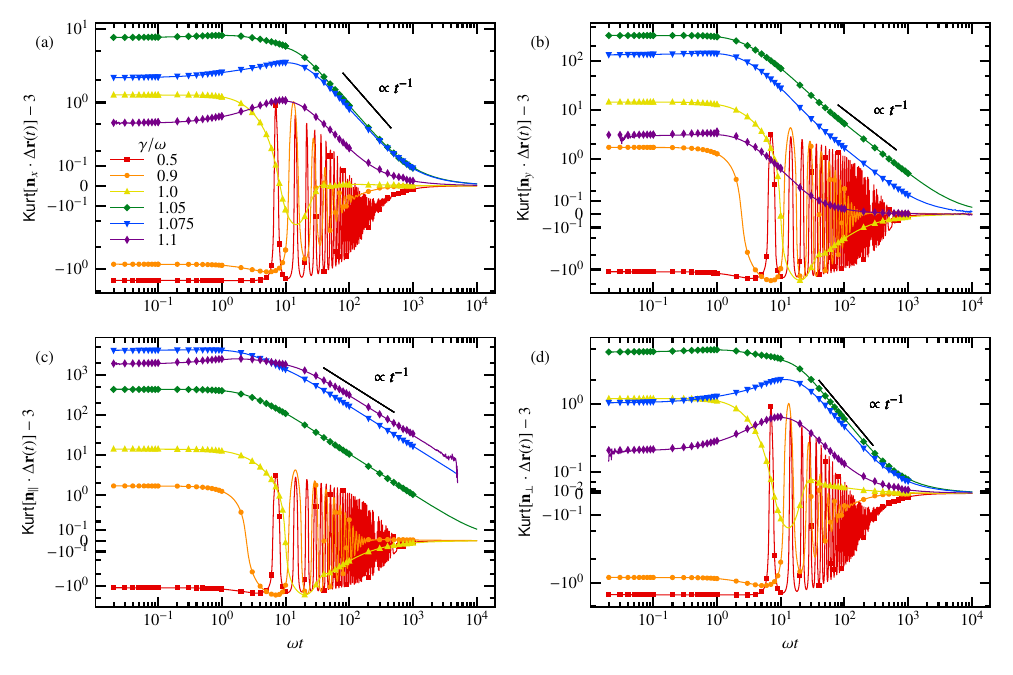}
	\caption{Rescaled kurtosis $\textsf{Kurt}[ \mathbf{n} \cdot \Delta  \mathbf{r}(t) ]-3$ for different values of the orienting torque $\gamma $ and four directions $ \mathbf{n}$ for a rotational diffusion coefficient $D_\text{rot}/\omega = 0.005$. The values are by a inverse-area-hyperbolic-sine function $y \to y_0 \sinh ^{-1}(y/y_0)$, with $y_0=0.15$. Full lines correspond to the spectral theory and symbols to the Langevin-simulation results. The black solid lines are auxiliary lines to serve as a guide to the eye.}
	\label{fig:Kurtosis_D_0.005}
\end{figure*}

\subsection{Skewness}
Similarly to the second cumulant, the third cumulant is obtained by comparing the third-order terms in \cref{eq:cumulant_expansion_3order_ISF,eq:cumulant_expansion_3order}. Rather than the third cumulant we discuss the skewness defined as 
\begin{align} \label{eq:skewness}
	\textsf{Skew}[\mathbf{n} \cdot \Delta \mathbf{r}(t)  ] &= \frac{\kappa_3[\mathbf{n} \cdot \Delta \mathbf{r}(t)]}{\kappa_2[\mathbf{n} \cdot \Delta \mathbf{r}(t)]^{3/2}} \nonumber \\
	&= \frac{\langle [ \mathbf{n} \cdot \Delta  \mathbf{r}(t) -\langle \mathbf{n} \cdot \Delta  \mathbf{r}(t) \rangle ]^3\rangle}{\langle [ \mathbf{n} \cdot \Delta  \mathbf{r}(t) -\langle \mathbf{n} \cdot \Delta  \mathbf{r}(t) \rangle ]^2\rangle ^{3/2}} .
\end{align}
Again, we analyze the skewness in \cref{fig:Skewness_D_0.005} for four different directions for a fixed diffusion coefficient and in each panel several values of the orienting torque $\gamma $ are shown. 
Oscillations between positive and negative values occur again in the running phase in all directions for times $t \gtrsim \tau_\omega$. 
In the locked phase, the skewness is purely negative (right-side skewed) for the  $x$, $y$, and parallel direction of observation, but with a difference in magnitude. In the perpendicular direction the skewness is positive (left-side skewed) and in $y$ and parallel direction the magnitude is much higher than in $x$ and perpendicular direction, which can be explained by the higher mean velocity in the $y$ and parallel direction for values around the bifurcation, see \cref{fig:mean_displacement_D_0_025}. 

The skewness is zero by construction in the harmonic approximation. We observe an increased skewness above and close to the bifurcation point, which are also values where the harmonic approximation does not hold well. The harmonic approximation holds if the barrier is sufficiently high, $\Delta U \gg k_\text{B} T$,  to ensure that Kramers' escape rate remains much lower than the harmonic relaxation rate~\cite{risken1996fokker,Chepizhko_2022}. 

For shorter times $t \lesssim \tau_\omega$ the skewness is constant with an enhanced value close but above the classical bifurcation.  
For long times $t \gtrsim \tau_\text{rot}$ every curve decays to zero since the angular motion is randomized and evolves to an effective diffusion. 

We find an algebraic tail in the decay of the form $\propto~t^{-1/2}$ or faster. This dynamics can be understood by assuming our random variables are the sum of identically and independently distributed increments
\begin{align}
	\Delta \mathbf{r} (N) = \sum_{n=1}^N \Delta \mathbf{r}_n,
\end{align}
since we can choose a time scale on which this assumption is true (coarse-grained dynamics). For such long times, the cumulants are extensive, $\kappa_j(N) \propto N $, and since the number of steps is proportional to time $t= N \Delta t$ it follows that $\kappa_j(t) \propto t $. 

Therefore, the skewness maximally decreases or increases for negative or positive skewness, respectively with 
\begin{align} 
	\frac{\kappa _3(t)}{\kappa _2 (t)^{3/2}} \propto \frac{t}{t^{3/2}} = t^{-1/2}.
\end{align}
This can also be applied for higher moments as the excess kurtosis, 
\begin{align} \label{eq:decayrate}
	\frac{\kappa _4(t)}{\kappa _2 (t)^{2} } \propto \frac{t}{t^{2}} = t^{-1},
\end{align}
which will be discussed in the following section. In conclusion, in accordance with the central limit theorem, we identified the speed of the convergence to a Gaussian distribution.

\subsection{Kurtosis}
Similar to the non-Gaussian parameter the kurtosis measures the degree of non-Gaussianity and is defined as 
\begin{align} \label{eq:kurtosis}
	\textsf{Kurt}[\mathbf{n} \cdot \Delta \mathbf{r}(t)  ]  &=  \frac{ \kappa_4[\mathbf{n} \cdot \Delta  \mathbf{r}(t)] }{\kappa_2[\mathbf{n} \cdot \Delta  \mathbf{r}(t)]^2 } + 3  \nonumber \\
	&=\frac{\langle [ \mathbf{n} \cdot \Delta  \mathbf{r}(t) -\langle \mathbf{n} \cdot \Delta  \mathbf{r}(t) \rangle ]^4\rangle }{\langle [ \mathbf{n} \cdot \Delta  \mathbf{r}(t) -\langle \mathbf{n} \cdot \Delta  \mathbf{r}(t) \rangle ]^2\rangle^2 },
\end{align}
where $\textsf{Kurt}[\mathbf{n} \cdot \Delta \mathbf{r}(t) ]-3 $ is called the excess kurtosis. As the expression for the fourth cumulant becomes lengthy it is convenient to decompose the fourth cumulant in its moments
\begin{align}
	\kappa_4 = m_4 -4 m_3m_1 - 3 m_2^2+12m_2m_1^2-6m_1^4,
\end{align} 
which can be computed from the characteristic function
\begin{align} \label{eq:moment_expansion}
	&F(\mathbf{k},t)= \sum_{j=0}^{\infty} \frac{(-\textsf{i} k)^j m_j[ \mathbf{n} \cdot \Delta \mathbf{r}(t) ]}{j!}= 1-  \textsf{i} k \textbf{n} \cdot \langle \Delta \textbf{r}(t) \rangle \nonumber \\ &  - \frac{k^2}{2} \langle [\textbf{n} \cdot \Delta \textbf{r}(t)]^2 \rangle  +  \frac{\textsf{i} k^3}{3!} \langle [\textbf{n} \cdot \Delta \textbf{r}(t)]^3 \rangle  + \frac{k^4}{4!} \langle [\textbf{n} \cdot \Delta \textbf{r}(t)]^4 \rangle \nonumber \\ &  + \cdots,
\end{align}
similarly to the cumulant-generating function, \cref{eq:cumulant_expansion_3order}. 

We then compare the expansion of the characteristic function to the moment expansion of the ISF, as shown in the \cref{sec:higher_moments}. The derivation is analogous to the derivation of the cumulant expansion of the ISF, \cref{eq:cumulant_expansion_3order_ISF}.
The final expression for the fourth moment is lengthy and therefore, it is deferred to the \cref{sec:higher_moments}.

We analyze the time evolution of the excess kurtosis again numerically and with Langevin simulations for various orienting torques $\gamma $ and in four different directions of observation, see \cref{fig:Kurtosis_D_0.005}. To visualize both wide positive and negative scales, we plot the excess kurtosis, $\textsf{Kurt}[\mathbf{n} \cdot \Delta \mathbf{r}(t) ]-3 $ using an inverse-area-hyperbolic-sine transformation $y \to y_0 \sinh ^{-1}(y/y_0)$, with $y_0=0.15$. The values around zero ($|y| \ll |y_0$|) are plotted in a linear scale, which is controlled by the parameter $y_0$.
In the linear range it is possible to show the oscillations between positive and negative values. 
For larger values $|y| \gg |y_0|$ the values can be interpreted similarly to a logarithmic scale, where values across multiple magnitudes and analytic tails can be visualized. This transformation allows us to plot both behaviors in a single graph.   

In all analyzed directions the excess kurtosis is constant for short times $t \lesssim \tau_\omega $ where this constant value is enhanced close to but still above the bifurcation $\gamma \gtrsim \omega $. For long times $t \gtrsim \tau_\text{rot}$ the excess kurtosis approaches zero, consistent with the expectation that the process becomes Gaussian and the angular motion becomes randomized and evolves into effective diffusion. Above the bifurcation, we observe a decay  $\propto t^{-1}$ or faster, see \cref{eq:decayrate}. Oscillations occur in the running phase $\gamma < \omega $ because of intrinsic drift of the particle, particularly noticeable at intermediate times $t \gtrsim \tau_\omega $.

The discrepancy between directions of observation lies in the magnitude of the excess kurtosis. Similarly to the skewness, in $x$ and the perpendicular direction  are much smaller in magnitude compared to the $y$ and parallel direction, especially above but close to the bifurcation. This is again explainable with the enhanced mean velocity in the $y$ and parallel direction, \cref{fig:mean_displacement_D_0_025}. In the harmonic approximation, the kurtosis is zero. Similar to the skewness, we observe increased kurtosis close and above the bifurcation point, where the harmonic approximation does not hold well.

The numerical values are much harder to obtain for the kurtosis than for lower-order cumulants and especially the extremes of very short and very long times were numerically challenging.  The explanation is that the matrices are truncated in the numerical process and this plays a major role in the precision of the measurements. 

\section{Conclusion}
We have investigated gravitaxis of a Brownian circle swimmer in two dimensions by deriving an analytical expression of the experimentally measurable ISF and expanded the ISF in its cumulants to obtain the mean velocity, variance, skewness, and kurtosis. The observables are analyzed in four directions of interest: parallel to the gravitational force $y$ and perpendicular to that in $x$-direction, as well as parallel and perpendicular to the mean velocity,  referred to as parallel and perpendicular direction. The solutions are obtained by a spectral-theoretical approach and validated by Langevin-dynamics simulations.

First, the model we use maps the particle's orientation to an overdamped noisy driven pendulum with a classical bifurcation when the orientational torque reaches the angular drift. This picture suggests an analysis of different orientational torques below (running phase), exactly at and above (locked phase) the classical bifurcation. We have interpreted our results in terms of a harmonic approximation, which is accurate, especially for a small diffusion coefficient and fixed torque above and not too close to the bifurcation.
In this regime the harmonic approximation is in good agreement with the numerically computed ISF and variance. We show the first eigenvalues and they coincide with the eigenvalues of the harmonic approximation within this regime and identified symmetries of the eigenvalues and eigenvectors. 

Second, we have analyzed the ISF for different orientational torques and wave vectors $\mathbf{k}v /\omega$ and demonstrated that transforming the ISF into the comoving frame eliminates oscillations caused by pure drift motion stemming from the non equilibrium dynamics. In the running phase, the dynamics are similar to the Brownian circle swimmer, with the difference that the imaginary part is highly oscillating, especially in the parallel direction . We observed persistent motion at large wavenumbers, circular motion at intermediate wavenumbers, and enhanced diffusion at small wavenumbers. Close to the bifurcation, oscillation frequency increase, with the fastest diffusion at small wavenumbers and persistent motion at intermediate wavenumbers.

In the locked phase, because of the mapping to the comoving frame, we found diffusion, that was magnitudes faster in perpendicular direction and the faster for higher wavenumbers. 

Additionally, we have investigated the variance and mean motion, as well as the skewness and kurtosis, for various directions and different gravitational torques.
In the running phase, the mean velocity in parallel direction is becoming significant close to the bifurcation as the swimmer oscillates for very low gravitational torques, i.e., does not move in a certain direction. In perpendicular direction the velocity is, by definition, zero.
The variance, skewness and kurtosis exhibit oscillations, at times comparable to the time it takes the particle to complete a circle. These oscillations are more pronounced the lower the gravitational torque and independent of the observation direction. 

In the locked phase, the mean velocity in the parallel direction reaches its maximum, while the variance is much lower compared to the perpendicular direction, where the mean velocity is zero. This behavior becomes exact in the harmonic approximation. In the variance we observe persistent, quadratic growth in time until the particle begins oscillating. For times when the particle has completed many oscillations, it exhibits diffusive behavior.
The skewness and kurtosis reach their highest values close but above the bifurcation, indicating big deviations from the harmonic approximation for those values. However, they converge to zero for high enough gravitational torques. They decay to zero for long times with a power law $\propto t^{-1/2}$ and $\propto t^{-1}$, respectively and thus indicating Gaussian behavior for times when the particle loses its orientation. 
In the parallel direction, the skewness is negative and large, while the kurtosis is positive and also relatively large. In contrast, in the perpendicular direction, both the skewness and kurtosis are positive but smaller compared to the parallel direction. These differences may be influenced by the mean velocity.

The model can be extented to the case of translational diffusion by adding noise and additional drift terms, as shown in the Supplemental Material of Ref.~\cite{Chepizhko_2022}. In our current formulation, the perturbing operator is linear in the wave vector, making the expansion of the ISF in orders of the wave vector straightforward. The translational diffusion generates also quadratic terms to the perturbing operator which makes the analysis slightly more involved without necessarily providing more insight in the rotational dynamics. As a further extension of the model, one can introduce angle-resolved spatial-temporal observables, which encode correlations between the initial and final angle and the displacement. This can be derived from the propagator, computed from the Fokker-Planck equation. A full characterization along these lines has been recently achieved for anisotropically diffusing colloidal dimers~\cite{Mayer_2021_dimers}. 

The observed effects should also be visible in the experiments with asymmetrical self-propelled particles of various origins. Bacteria moving in circles, such as \emph{E. coli} and \emph{V. cholerae} can serve as examples from the biological world. Various asymmetrical artificial microswimmers have also been developed in recent years.
We are not limited to gravity as a means to produce external torque. Other forces,  such as hydrodynamic ones in a laminar flow or magnetic ones can be suitable too for experimental realizations. The interactions of many chiral active particles are another avenue where the presence of the external torque can be important to investigate. The collective intermediate scattering function can be explored based on the individual one studied here. Extending the model to three dimensions would make it even more comparable to the real world as well as the expansion to a asymmetric colloid in a viscoelastic fluid, similar to \cite{narinder2018memory}, as they also found a transition from angular diffusion to persistent rotational motion. Complex environments, such as obstacles in assymetrically pattered arrays \cite{reichhardt2013dynamics} or patterned environment \cite{peng2016controlling} are interesting systems to address in the future.

\begin{acknowledgements}
We thank Anton L\"uders and Christina Kurzthaler for helpful discussions and acknowledge the use of AI for its assistance with grammar checking, translations, and text enhancement. 
OC is supported by the Austrian Science Fund (FWF): M 2450-NBL. TF acknowledges
funding by the Austrian Science Fund (FWF): P 35580-N.
\end{acknowledgements}
\begin{widetext}

	\appendix
\section{Symmetries of the operators}\label{sec:matrix_elements}
The matrix elements \cref{eq:matrix_representation_L} have the symmetry  $\langle (-\beta) | \mathcal{L} (-\alpha) \rangle ^* = \langle \beta | \mathcal{L}  \alpha \rangle$ resulting in a specific symmetry of the eigenvalues and eigenvectors which we describe in the following. Rewriting the eigenvalue equation for the right eigenvector yields
\begin{align} 
	 \sum_{\alpha} \mathcal{L}_{\beta \alpha} \langle \alpha |r_n \rangle &= \lambda_{n} \langle \beta | r_n \rangle \nonumber  \Leftrightarrow \\
	 \sum_{\alpha} \mathcal{L}_{(-\beta) (-\alpha)} \langle -\alpha |r_n \rangle &= \lambda_{n} \langle -\beta | r_n \rangle \nonumber  \Leftrightarrow \\
	 \sum_{\alpha} \mathcal{L}_{\beta\alpha}^* \langle -\alpha |r_n \rangle &= \lambda_{n} \langle -\beta | r_n \rangle \nonumber  \Leftrightarrow \\
	 \sum_{\alpha} \mathcal{L}_{\beta\alpha} \langle -\alpha |r_n \rangle^* &= \lambda_{n}^* \langle -\beta | r_n \rangle^* \nonumber  ,
	\end{align}
and we can identify for each eigenvector $|r_{n} \rangle = \sum\nolimits_\alpha |\alpha \rangle \langle \alpha | r_n \rangle$ with eigenvalue $\lambda_{n}$ an eigenvector $|r_{-n} \rangle \coloneq  \sum\nolimits_\alpha |\alpha \rangle \langle -\alpha| r_n \rangle ^*$ with eigenvalue $\lambda_{-n} \coloneq  \lambda^*_{n}$. A similar relation   can be shown for the left eigenvector.

Moreover, the tridiagonal matrix $\mathcal{L}_{\beta \alpha }$ has a zero line in the middle, $\mathcal{L}_{0 \alpha}=0$, and each right eigenvector $n\neq 0$ has a zero entry in the middle, $\langle 0 | r_n \rangle =0 $. This follows from the orthogonality, $\langle l_0 | r_n \rangle =1 $, and the fact that $\langle l_0 | = \langle 0 |  \Leftrightarrow  \langle l_0 | 0 \rangle =1$.
Therefore, in the eigenvalue equation, the upper left or lower right block of the matrix do not couple to each other. Consequently, the eigenvectors $| r_n \rangle $ have entries exclusively in either the upper half,  $\langle \alpha | r_n \rangle = 0 $ for $n>0$ and $\alpha \leq 0 $ or lower half, $n<0$ and $\alpha \geq 0 $, respectively. 
As in  $(\mathcal{L}^\dagger)_{\beta \alpha}$ there is a zero column, directly following from the zero line in $\mathcal{L}$, this also holds for the left eigenvectors. Similarly, we find that the eigenvectors $\langle l_n |$ have entries exclusively in either the upper or lower half, including zero, and we find $\langle l_n| \alpha  \rangle = 0 $ for $n>0$ and $\alpha < 0 $ or, $n<0$ and $\alpha > 0 $, respectively. 

Furthermore, the operator $ \delta  \mathcal{L}_{\mathbf{k}} $ has the symmetry $\langle (-\beta) |\delta \mathcal{L}_{\mathbf{k}}  (-\alpha) \rangle ^* = - \langle \beta |\delta \mathcal{L}_{\mathbf{k}}  \alpha \rangle$ and we conclude a symmetry for the matrix elements which are used to compute the cumulants and moments,
\begin{align} \label{eq:matrix_elements_symmetry}
	\langle l_n |  \delta  \mathcal{L}_{\mathbf{k}}  r_m \rangle^* &=\sum_{\alpha, \beta}  \langle l_n | \alpha \rangle^*  \langle \alpha |  \delta  \mathcal{L}_{\mathbf{k}}   | \beta \rangle ^* \langle  \beta  |r_m \rangle^* \nonumber \\
	&=-\sum_{\alpha, \beta}  \langle l_n | \alpha \rangle^*  \langle (-\alpha) |\delta  \mathcal{L}_{\mathbf{k}} (-\beta) \rangle  \langle  \beta  |r_m \rangle^*  \nonumber \\
	&=-\sum_{\alpha, \beta}  \langle l_n | (-\alpha) \rangle^*  \langle \alpha |\delta  \mathcal{L}_{\mathbf{k}} \beta \rangle  \langle  (-\beta)  |r_m \rangle^* \nonumber \\
	&= - \langle l_{-n} |  \delta  \mathcal{L}_{\mathbf{k}}  r_{-m} \rangle  .
\end{align}

Considering this symmetry, 
 we conclude that the variance is real 
\begin{align}
	&\textsf{Var} [\mathbf{n} \cdot \Delta \mathbf{r}(t)  ] = - \frac{2}{k^2} \sum_{n \neq 0} \frac{e^{-\lambda_n t} + \lambda_n t -1}{\lambda_n^2} 
	\langle l_0 |\delta \mathcal{L}_{\mathbf{k}}  r_n \rangle \langle l_n |  \delta\mathcal{L}_{\mathbf{k}}   r_0 \rangle = - \frac{4}{k^2} \sum_{n >0} \text{Re}\left[ \frac{e^{-\lambda_n t} + \lambda_n t -1}{\lambda_n^2} 
	\langle l_0 |\delta \mathcal{L}_{\mathbf{k}}  r_n \rangle \langle l_n |  \delta\mathcal{L}_{\mathbf{k}}   r_0 \rangle \right]  .
\end{align}
It can be checked that also higher cumulants are real because of these symmetries. The minus sign in the last line of \cref{eq:matrix_elements_symmetry} determines whether the real or imaginary part contributes to the  cumulant expression. For even orders of $k$, the real part contributes, while for odd orders of $k$, the imaginary part contributes.

\section{Time evolution operator} \label{sec:time_evolution}
	To expand the ISF \cref{eq:ISF2} we iteratively substitute the time evolution operator with the Dyson representation \cref{eq:dyson_representation}. The time evolution operator up to the fourth order then reads
	\begin{align} \label{eq:dyson_representation_iterative}
		e^{(\mathcal{L}+\delta \mathcal{L}_{\mathbf{k}} ) t } =&e^{\mathcal{L} t} + 
			\int_0^t \!\diff s \, e^{\mathcal{L} (t-s)} \delta \mathcal{L}_{\mathbf{k}} \left( e^{\mathcal{L} s} + \int_0^s \!\diff u \, e^{\mathcal{L} (s-u)} \delta \mathcal{L}_{\mathbf{k}} \left( e^{\mathcal{L} u} + \int_0^u \!\diff w \, e^{\mathcal{L} (u-w)} \delta \mathcal{L}_{\mathbf{k}} \left( e^{\mathcal{L} w} + \int_0^w \!\diff r \, e^{\mathcal{L} (w-r)} \delta \mathcal{L}_{\mathbf{k}} e^{(\mathcal{L}+\delta\mathcal{L}_{\mathbf{k}} ) r }  \right)   \right)  \right)   \nonumber \\
			=&e^{\mathcal{L} t} + \int_0^t \!\diff s \, e^{\mathcal{L} (t-s)} \delta \mathcal{L}_{\mathbf{k}} e^{\mathcal{L} s} + \int_0^t \!\diff s \int_0^s \!\diff u \, e^{\mathcal{L} (t-s)} \delta  \mathcal{L}_{\mathbf{k}}  e^{\mathcal{L} (s-u)} \delta \mathcal{L}_{\mathbf{k}} e^{\mathcal{L} u} +  \nonumber  \\
			&+ \int_0^t \!\diff s  \int_0^s \!\diff u \int_0^u \!\diff w \, e^{\mathcal{L} (t-s)} \delta \mathcal{L}_{\mathbf{k}} e^{\mathcal{L} (s-u)} \delta \mathcal{L}_{\mathbf{k}} e^{\mathcal{L} (u-w)} \delta \mathcal{L}_{\mathbf{k}}  e^{\mathcal{L} w} +   \nonumber \\
			&+ \int_0^t\!\diff s \int_0^s \!\diff u \int_0^u \!\diff w  \int_0^w \!\diff r\, e^{\mathcal{L} (t-s)} \delta \mathcal{L}_{\mathbf{k}}  e^{\mathcal{L} (s-u)} \delta \mathcal{L}_{\mathbf{k}}  e^{\mathcal{L} (u-w)} \delta \mathcal{L}_{\mathbf{k}}   e^{\mathcal{L} (w-r)} \delta \mathcal{L}_{\mathbf{k}} e^{\mathcal{L}  r }  +  O(\delta \mathcal{L}_{\mathbf{k}})^5 .
	\end{align} 

 \section{DDM  corrected with drift} \label{sec:DDM}

 Differential dynamic microscopy (DDM) has been first introduced by Cerbino and Trappe~\cite{Cerbino_2008} to measure the ISF experimentally, relying only on advanced imaging techniques, rather than scattering experiments. The theoretical background of DDM is nicely presented in Refs.~\cite{Giavazzi_2009, Wilson_2011}. 
 This technique can also be adapted to the comoving frame, indicated in the literature as well~\cite{edera2021deformation, giavazzi2014digital, richards2021particle}.
 In the DDM method, a sequence of microscopic images is taken, and dynamic contrast is measured. In the comoving frame, the image is shifted with velocity $v$ to include a drift and so the difference of the two images 
recorded at time $t$ and $0$ 
is 
 \begin{align}
	\Delta I (\mathbf{r}, t ) =  I (\mathbf{r}, t ) -  I (\mathbf{r}- \mathbf{v} t, 0 ).
 \end{align} 
Here, the time- and space-dependent intensity 
\begin{align}
	I(\mathbf{r},t)= \int \kappa (\mathbf{r}-\mathbf{r}^\prime) c(\mathbf{r}^\prime,t) \diff \mathbf{r}^\prime +\eta (\mathbf{r},t) ,
\end{align} 
is a function of the concentration, $c(\mathbf{r},t)= \sum_{i=1}^N \delta (\mathbf{r}-\mathbf{R}_i(t) )$, the point spread function, $\kappa (\mathbf{r}) $, and the camera noise, $\eta (\mathbf{r},t)$, which is assumed to be uncorrelated with the signal. 
The measurable quantity and DDM signal is $\Delta D(\mathbf{k}, t )  =\langle |\Delta I(\mathbf{k},t)|^2 \rangle $, where the Fourier transformation of the intensity is 
\begin{align}
	\Delta I(\mathbf{k}, t ) = \kappa(\mathbf{k})[c(\mathbf{k},t)-e^{-i \mathbf{k} \cdot \mathbf{v} t }c(\mathbf{k},0)]+ \Delta \eta (\mathbf{k},t).
\end{align}
The static structure factor and ISF are computed by the concentration 
\begin{align}
  F(\mathbf{k}) = \frac{1}{N} \langle |c (\mathbf{k})|^2 \rangle, \qquad  F(\mathbf{k},t) = \frac{1}{N} \langle c (\mathbf{k},t)  c (\mathbf{k},0)^*\rangle.
\end{align}
Using the relation, $\int \diff \mathbf{r}  e^{-i \mathbf{k} \cdot \mathbf{r} t } I (\mathbf{r}-\mathbf{v}t, 0 ) = e^{-i \mathbf{k} \cdot \mathbf{v} t } I(\mathbf{k}, t )$, we obtain the final expression for the signal
\begin{align}
	&\Delta D (\mathbf{k}, t ) =  a(\mathbf{k}) \left(1- \text{Re}\left[e^{i \mathbf{k} \cdot \mathbf{v}t } \frac{F(\mathbf{k},t)}{F(\mathbf{k})}\right]\right)+ b(\mathbf{k}),
\end{align}
which connects the signal to the ISF in the comoving frame.
We identify the additional term $b(\mathbf{k})= \langle | \Delta \eta (\mathbf{k})|^2 \rangle $ as the detection noise, and the term $a(\mathbf{k})  = 2 N |\kappa (\mathbf{k})|^2 F(\mathbf{k})$ as a static amplitude term that depends on the point spread function of the detector and structure factor.

\section{Fourth-order moment} \label{sec:higher_moments}

To find the fourth moment  $m_4$ we expand the ISF in moments, similarly to the cumulant expansion \cref{eq:cumulant_expansion_3order_ISF} up to the fourth order 
	\begin{align} \label{eq:ISF3}
		&F(\mathbf{k},t) 	 = 1  +  t \langle l_0 |   \delta \mathcal{L}_{\mathbf{k}}  r_0\rangle + 
		\sum_{n } \frac{e^{-\lambda_n t} + \lambda_n t -1}{\lambda_n^2} 
		\langle l_0 |\delta \mathcal{L}_{\mathbf{k}}  r_n \rangle \langle l_n |  \delta\mathcal{L}_{\mathbf{k}}   r_0 \rangle \nonumber
		\\ 
		&+\sum_{n} \sum_{m}  \left(\frac{e^{-\lambda_n t}+\lambda_n  t-1}{\lambda_n ^2 (\lambda_m -\lambda_n )}  + \frac{e^{-\lambda_m t}+\lambda_m  t-1}{\lambda_m ^2 (\lambda_n -\lambda_m )}\right) \langle l_0 | \delta \mathcal{L}_{\mathbf{k}}   r_n \rangle \langle l_n | \delta \mathcal{L}_{\mathbf{k}}    r_m \rangle \langle l_m | \delta \mathcal{L}_{\mathbf{k}}  r_0 \rangle  \nonumber \\
		&+\sum_{n} \sum_{m} \sum_{p} 
		\left(\frac{e^{-\lambda_n t}+\lambda_n  t-1}{\lambda_n ^2 (\lambda_n -\lambda_m ) (\lambda_n -\lambda_p )}+\frac{e^{-\lambda_m t}+\lambda_m  t-1}{\lambda_m ^2 (\lambda_m -\lambda_n ) (\lambda_m -\lambda_o )}+\frac{e^{-\lambda_p t}+\lambda_p  t-1}{\lambda_p ^2 (\lambda_p -\lambda_m ) (\lambda_p -\lambda_n ) } \right) \nonumber \\
		&\times \langle l_0 | \delta \mathcal{L}_{\mathbf{k}}     r_n \rangle \langle l_n | \delta \mathcal{L}_{\mathbf{k}}  r_m \rangle \langle l_m | \delta \mathcal{L}_{\mathbf{k}}     r_p \rangle \langle l_p | \delta \mathcal{L}_{\mathbf{k}} r_0 \rangle + O(|\mathbf{k}|^5).
	\end{align}
Here, all sums run over all integers. 
This then is compared to the expansion in  moments \cref{eq:moment_expansion} to identify the fourth moment. This expression formally contains divisions by zero, they arise since we should have distinguished the cases of exponentials and constants before performing the integrals. However, this can be easily remedied since the terms can be analytically continued. 
The result for the fourth moment then is 

\begin{equation}
	\begin{aligned} \label{eq:4thmoment}
	&\frac{1}{4!} \langle [\textbf{k} \cdot \Delta \textbf{r}(t)]^4 \rangle  =  \\
	&\sum_n \sum_m \sum_p 
	\left(\frac{e^{-\lambda_n t}+\lambda_n  t-1}{\lambda_n ^2 (\lambda_n -\lambda_m ) (\lambda_n -\lambda_p )}+\frac{e^{-\lambda_m t}+\lambda_m  t-1}{\lambda_m ^2 (\lambda_m -\lambda_n ) (\lambda_m -\lambda_p )}+\frac{e^{-\lambda_p t}+\lambda_p  t-1}{\lambda_p ^2 (\lambda_p -\lambda_m ) (\lambda_p -\lambda_n ) } \right) \\
	& \times	\langle l_0 | \delta \mathcal{L}_{\mathbf{k}}    r_n \rangle \langle l_n | \delta \mathcal{L}_{\mathbf{k}}   r_m \rangle \langle l_m | \delta \mathcal{L}_{\mathbf{k}}     r_p \rangle \langle l_p | \delta \mathcal{L}_{\mathbf{k}} r_0 \rangle =\frac{t^4}{24} \langle l_0 | \delta \mathcal{L}_{\mathbf{k}} r_0 \rangle^4 \\
	&+ 3\sum _{n \neq 0} \frac{\lambda_n  t (\lambda_n  t (\lambda_n  t-3)+6)+6 e^{\lambda_n  (-t)}-6}{6 \lambda_n ^4} \langle l_0 | \delta \mathcal{L}_{\mathbf{k}}     r_n \rangle \langle l_n | \delta \mathcal{L}_{\mathbf{k}}   r_0 \rangle \langle l_0 | \delta \mathcal{L}_{\mathbf{k}}      r_0 \rangle \langle l_0 | \delta \mathcal{L}_{\mathbf{k}} r_0 \rangle  \\
	&+2\sum _{n \neq 0}
	\frac{\lambda_n ^2 t^2-4 \lambda_n  t-2 e^{\lambda_n  (-t)} (\lambda_n  t+3)+6}{2 \lambda_n ^4}
	\langle l_0 | \delta \mathcal{L}_{\mathbf{k}}     r_n \rangle \langle l_n| \delta \mathcal{L}_{\mathbf{k}}   r_n \rangle \langle l_n | \delta \mathcal{L}_{\mathbf{k}}      r_0 \rangle \langle l_0 | \delta \mathcal{L}_{\mathbf{k}}r_0 \rangle \\
	&+2\sum _{n \neq 0} \sum _{m \neq 0, m \neq n}  \left(
	\frac{t^2}{2 \lambda_m  \lambda_n }+\frac{\lambda_m  t+e^{\lambda_m  (-t)}-1}{\lambda_m ^3 (\lambda_m -\lambda_n )}+\frac{\lambda_n  t+e^{\lambda_n  (-t)}-1}{\lambda_n ^3 (\lambda_n -\lambda_m )} \right)
	\langle l_0 | \delta \mathcal{L}_{\mathbf{k}}     r_n \rangle \langle l_n | \delta \mathcal{L}_{\mathbf{k}}   r_m \rangle \langle l_m | \delta \mathcal{L}_{\mathbf{k}}    r_0 \rangle \langle l_0 | \delta \mathcal{L}_{\mathbf{k}} r_0 \rangle \\
	& +\sum _{n \neq 0}  
	\frac{\lambda_n ^2 t^2-4 \lambda_n  t-2 e^{\lambda_n  (-t)} (\lambda_n  t+3)+6}{2 \lambda_n ^4}
	\langle l_0 | \delta \mathcal{L}_{\mathbf{k}}     r_n \rangle \langle l_n | \delta \mathcal{L}_{\mathbf{k}}   r_0 \rangle \langle l_0 | \delta \mathcal{L}_{\mathbf{k}}     r_n \rangle \langle l_n | \delta \mathcal{L}_{\mathbf{k}} r_0 \rangle \\
	&+\sum _{n \neq 0} \sum _{m \neq 0,m \neq n} \left(  \frac{t^2}{2 \lambda_m  \lambda_n }+\frac{\lambda_m  t+e^{\lambda_m  (-t)}-1}{\lambda_m ^3 (\lambda_m -\lambda_n )}+\frac{\lambda_n  t+e^{\lambda_n  (-t)}-1}{\lambda_n ^3 (\lambda_n -\lambda_m )} \right)
	\langle l_0 | \delta \mathcal{L}_{\mathbf{k}}     r_n \rangle \langle l_n | \delta \mathcal{L}_{\mathbf{k}}  r_0 \rangle \langle l_0 | \delta \mathcal{L}_{\mathbf{k}}     r_m \rangle \langle l_m | \delta \mathcal{L}_{\mathbf{k}} r_0 \rangle \\
	&+\sum _{n \neq 0} 
	\frac{e^{\lambda_n  (-t)} \left(\lambda_n ^2 t^2+4 \lambda_n  t+2 e^{\lambda_n  t} (\lambda_n  t-3)+6\right)}{2 \lambda_n ^4} 
	\langle l_0 | \delta \mathcal{L}_{\mathbf{k}}     r_n \rangle \langle l_n | \delta \mathcal{L}_{\mathbf{k}}   r_n \rangle \langle l_n | \delta \mathcal{L}_{\mathbf{k}}      r_n \rangle \langle l_n | \delta \mathcal{L}_{\mathbf{k}} r_0 \rangle\\
	&+\sum _{n \neq 0} \sum _{m \neq 0,m \neq n} 
	\frac{e^{-t (\lambda_m +\lambda_n )} \left(\lambda_m ^3 e^{\lambda_m  t}+\lambda_n ^2 e^{\lambda_n  t} \left(2 \lambda_n +\lambda_m ^2 (-t)+\lambda_m  (\lambda_n  t-3)\right)+(\lambda_m -\lambda_n )^2 e^{t (\lambda_m +\lambda_n )} (\lambda_m  (\lambda_n  t-1)-2 \lambda_n )\right)}{\lambda_m ^3 \lambda_n ^2 (\lambda_m -\lambda_n )^2} \\
	& \times \langle l_0 | \delta \mathcal{L}_{\mathbf{k}}    r_n \rangle \langle l_n | \delta \mathcal{L}_{\mathbf{k}}   r_m \rangle \langle l_m | \delta \mathcal{L}_{\mathbf{k}}     r_p \rangle \langle l_p | \delta \mathcal{L}_{\mathbf{k}} r_0 \rangle \\
	&+\sum _{n \neq 0} \sum _{m \neq 0,m \neq n} 
	\frac{e^{-t (\lambda_m +\lambda_n )} \left(\lambda_m ^2 e^{\lambda_m  t} (\lambda_m  (\lambda_n  t+2)-\lambda_n  (\lambda_n  t+3))+(\lambda_m -\lambda_n )^2 e^{t (\lambda_m +\lambda_n )} (\lambda_m  (\lambda_n  t-2)-\lambda_n )+\lambda_n ^3 e^{\lambda_n  t}\right)}{\lambda_m ^2 \lambda_n ^3 (\lambda_m -\lambda_n )^2}\\
	& \times \langle l_0 | \delta \mathcal{L}_{\mathbf{k}}     r_n \rangle \langle l_n | \delta \mathcal{L}_{\mathbf{k}}   r_m \rangle \langle l_m | \delta \mathcal{L}_{\mathbf{k}}     r_p \rangle \langle l_p | \delta \mathcal{L}_{\mathbf{k}} r_0 \rangle  \\
	&+\sum _{n \neq 0} \sum _{m \neq 0,m \neq n} 
	\frac{e^{-t (\lambda_m +\lambda_n )} \left(\lambda_m ^2 e^{\lambda_m  t} (\lambda_m  (\lambda_n  t+2)-\lambda_n  (\lambda_n  t+3))+(\lambda_m -\lambda_n )^2 e^{t (\lambda_m +\lambda_n )} (\lambda_m  (\lambda_n  t-2)-\lambda_n )+\lambda_n ^3 e^{\lambda_n  t}\right)}{\lambda_m ^2 \lambda_n ^3 (\lambda_m -\lambda_n )^2}\\
	&	\times \langle l_0 | \delta \mathcal{L}_{\mathbf{k}}     r_n \rangle \langle l_n | \delta \mathcal{L}_{\mathbf{k}}   r_m \rangle \langle l_m | \delta \mathcal{L}_{\mathbf{k}}     r_p \rangle \langle l_p | \delta \mathcal{L}_{\mathbf{k}} r_0 \rangle \\
	&+\sum _{n \neq 0} \sum _{m \neq 0,m \neq n} \sum _{p \neq 0,p \neq m,p \neq n} 
	\frac{\lambda_m  t+e^{\lambda_m  (-t)}-1}{\lambda_m ^2 (\lambda_m -\lambda_n ) (\lambda_m -\lambda_p )}+\frac{\lambda_n  t+e^{\lambda_n  (-t)}-1}{\lambda_n ^2 (\lambda_n -\lambda_m ) (\lambda_n -\lambda_p )}+\frac{\lambda_p  t+e^{\lambda_p  (-t)}-1}{\lambda_p ^2 (\lambda_p -\lambda_m ) (\lambda_p -\lambda_n )}\\
	&	\times \langle l_0 | \delta \mathcal{L}_{\mathbf{k}}     r_n \rangle \langle l_n | \delta \mathcal{L}_{\mathbf{k}}  r_m \rangle \langle l_m | \delta \mathcal{L}_{\mathbf{k}}     r_p \rangle \langle l_p | \delta \mathcal{L}_{\mathbf{k}} r_0 \rangle .
\end{aligned}
\end{equation}

\end{widetext}


\begin{thebibliography}{47}%
	\makeatletter
	\providecommand \@ifxundefined [1]{%
	 \@ifx{#1\undefined}
	}%
	\providecommand \@ifnum [1]{%
	 \ifnum #1\expandafter \@firstoftwo
	 \else \expandafter \@secondoftwo
	 \fi
	}%
	\providecommand \@ifx [1]{%
	 \ifx #1\expandafter \@firstoftwo
	 \else \expandafter \@secondoftwo
	 \fi
	}%
	\providecommand \natexlab [1]{#1}%
	\providecommand \enquote  [1]{``#1''}%
	\providecommand \bibnamefont  [1]{#1}%
	\providecommand \bibfnamefont [1]{#1}%
	\providecommand \citenamefont [1]{#1}%
	\providecommand \href@noop [0]{\@secondoftwo}%
	\providecommand \href[0]{\begingroup \@sanitize@url \@href}%
	\providecommand \@href[1]{\@@startlink{#1}\@@href}%
	\providecommand \@@href[1]{\endgroup#1\@@endlink}%
	\providecommand \@sanitize@url [0]{\catcode `\\12\catcode `\$12\catcode `\&12\catcode `\#12\catcode `\^12\catcode `\_12\catcode `\%12\relax}%
	\providecommand \@@startlink[1]{}%
	\providecommand \@@endlink[0]{}%
	\providecommand \url  [0]{\begingroup\@sanitize@url \@url }%
	\providecommand \@url [1]{\endgroup\@href {#1}{\urlprefix }}%
	\providecommand \urlprefix  [0]{URL }%
	\providecommand \Eprint [0]{\href }%
	\providecommand \doibase [0]{http://dx.doi.org/}%
	\providecommand \selectlanguage [0]{\@gobble}%
	\providecommand \bibinfo  [0]{\@secondoftwo}%
	\providecommand \bibfield  [0]{\@secondoftwo}%
	\providecommand \translation [1]{[#1]}%
	\providecommand \BibitemOpen [0]{}%
	\providecommand \bibitemStop [0]{}%
	\providecommand \bibitemNoStop [0]{.\EOS\space}%
	\providecommand \EOS [0]{\spacefactor3000\relax}%
	\providecommand \BibitemShut  [1]{\csname bibitem#1\endcsname}%
	\let\auto@bib@innerbib\@empty
	\bibitem [{\citenamefont {Nosrati}\ \emph {et~al.}(2015)\citenamefont {Nosrati}, \citenamefont {Driouchi}, \citenamefont {Yip},\ and\ \citenamefont {Sinton}}]{Nosrati_2015}%
	  \BibitemOpen
	  \bibfield  {author} {\bibinfo {author} {\bibfnamefont {R.}~\bibnamefont {Nosrati}}, \bibinfo {author} {\bibfnamefont {A.}~\bibnamefont {Driouchi}}, \bibinfo {author} {\bibfnamefont {C.}~\bibnamefont {Yip}}, \ and\ \bibinfo {author} {\bibfnamefont {D.}~\bibnamefont {Sinton}},\ }\bibfield  {title} {\enquote {\bibinfo {title} {Two-dimensional slither swimming of sperm within a micrometre of a surface},}\ }\href{\doibase 10.1038/ncomms9703} {\bibfield  {journal} {\bibinfo  {journal} {Nature communications}\ }\textbf {\bibinfo {volume} {6}},\ \bibinfo {pages} {8703} (\bibinfo {year} {2015})}\BibitemShut {NoStop}%
	\bibitem [{\citenamefont {Schwarz}\ \emph {et~al.}(2020)\citenamefont {Schwarz}, \citenamefont {Medina-S{\'a}nchez},\ and\ \citenamefont {Schmidt}}]{schwarz2020sperm}%
	  \BibitemOpen
	  \bibfield  {author} {\bibinfo {author} {\bibfnamefont {L.}~\bibnamefont {Schwarz}}, \bibinfo {author} {\bibfnamefont {M.}~\bibnamefont {Medina-S{\'a}nchez}}, \ and\ \bibinfo {author} {\bibfnamefont {O.~G.}\ \bibnamefont {Schmidt}},\ }\bibfield  {title} {\enquote {\bibinfo {title} {Sperm-hybrid micromotors: on-board assistance for nature’s bustling swimmers},}\ }\href{\doibase https://doi.org/10.1530/rep-19-0096} {\bibfield  {journal} {\bibinfo  {journal} {Reproduction}\ }\textbf {\bibinfo {volume} {159}},\ \bibinfo {pages} {R83} (\bibinfo {year} {2020})}\BibitemShut {NoStop}%
	\bibitem [{\citenamefont {Debnath}\ \emph {et~al.}(2020)\citenamefont {Debnath}, \citenamefont {Ghosh}, \citenamefont {Misko}, \citenamefont {Li}, \citenamefont {Marchesoni},\ and\ \citenamefont {Nori}}]{debnath2020enhanced}%
	  \BibitemOpen
	  \bibfield  {author} {\bibinfo {author} {\bibfnamefont {D.}~\bibnamefont {Debnath}}, \bibinfo {author} {\bibfnamefont {P.~K.}\ \bibnamefont {Ghosh}}, \bibinfo {author} {\bibfnamefont {V.~R.}\ \bibnamefont {Misko}}, \bibinfo {author} {\bibfnamefont {Y.}~\bibnamefont {Li}}, \bibinfo {author} {\bibfnamefont {F.}~\bibnamefont {Marchesoni}}, \ and\ \bibinfo {author} {\bibfnamefont {F.}~\bibnamefont {Nori}},\ }\bibfield  {title} {\enquote {\bibinfo {title} {Enhanced motility in a binary mixture of active nano/microswimmers},}\ }\href{\doibase http://xlink.rsc.org/?DOI=d0nr01765e} {\bibfield  {journal} {\bibinfo  {journal} {Nanoscale}\ }\textbf {\bibinfo {volume} {12}},\ \bibinfo {pages} {9717} (\bibinfo {year} {2020})}\BibitemShut {NoStop}%
	\bibitem [{\citenamefont {Baraban}\ \emph {et~al.}(2012)\citenamefont {Baraban}, \citenamefont {Tasinkevych}, \citenamefont {Popescu}, \citenamefont {Sanchez}, \citenamefont {Dietrich},\ and\ \citenamefont {Schmidt}}]{Baraban_2012}%
	  \BibitemOpen
	  \bibfield  {author} {\bibinfo {author} {\bibfnamefont {L.}~\bibnamefont {Baraban}}, \bibinfo {author} {\bibfnamefont {M.}~\bibnamefont {Tasinkevych}}, \bibinfo {author} {\bibfnamefont {M.~N.}\ \bibnamefont {Popescu}}, \bibinfo {author} {\bibfnamefont {S.}~\bibnamefont {Sanchez}}, \bibinfo {author} {\bibfnamefont {S.}~\bibnamefont {Dietrich}}, \ and\ \bibinfo {author} {\bibfnamefont {O.~G.}\ \bibnamefont {Schmidt}},\ }\bibfield  {title} {\enquote {\bibinfo {title} {Transport of cargo by catalytic {J}anus micro-motors},}\ }\href{\doibase 10.1039/C1SM06512B} {\bibfield  {journal} {\bibinfo  {journal} {Soft Matter}\ }\textbf {\bibinfo {volume} {8}},\ \bibinfo {pages} {48} (\bibinfo {year} {2012})}\BibitemShut {NoStop}%
	\bibitem [{\citenamefont {Garc\'{i}a}\ \emph {et~al.}(2013)\citenamefont {Garc\'{i}a}, \citenamefont {Orozco}, \citenamefont {Guix}, \citenamefont {Gao}, \citenamefont {Sattayasamitsathit}, \citenamefont {Escarpa}, \citenamefont {Merkoçi},\ and\ \citenamefont {Wang}}]{Jahir_2013}%
	  \BibitemOpen
	  \bibfield  {author} {\bibinfo {author} {\bibfnamefont {M.}~\bibnamefont {Garc\'{i}a}}, \bibinfo {author} {\bibfnamefont {J.}~\bibnamefont {Orozco}}, \bibinfo {author} {\bibfnamefont {M.}~\bibnamefont {Guix}}, \bibinfo {author} {\bibfnamefont {W.}~\bibnamefont {Gao}}, \bibinfo {author} {\bibfnamefont {S.}~\bibnamefont {Sattayasamitsathit}}, \bibinfo {author} {\bibfnamefont {A.}~\bibnamefont {Escarpa}}, \bibinfo {author} {\bibfnamefont {A.}~\bibnamefont {Merkoçi}}, \ and\ \bibinfo {author} {\bibfnamefont {J.}~\bibnamefont {Wang}},\ }\bibfield  {title} {\enquote {\bibinfo {title} {Micromotor-based lab-on-chip immunoassays},}\ }\href{\doibase 10.1039/C2NR32400H} {\bibfield  {journal} {\bibinfo  {journal} {Nanoscale}\ }\textbf {\bibinfo {volume} {5}},\ \bibinfo {pages} {1325} (\bibinfo {year} {2013})}\BibitemShut {NoStop}%
	\bibitem [{\citenamefont {Koumakis}\ \emph {et~al.}(2013)\citenamefont {Koumakis}, \citenamefont {Lepore}, \citenamefont {Maggi},\ and\ \citenamefont {Leonardo}}]{Koumakis_2013}%
	  \BibitemOpen
	  \bibfield  {author} {\bibinfo {author} {\bibfnamefont {N.}~\bibnamefont {Koumakis}}, \bibinfo {author} {\bibfnamefont {A.}~\bibnamefont {Lepore}}, \bibinfo {author} {\bibfnamefont {C.}~\bibnamefont {Maggi}}, \ and\ \bibinfo {author} {\bibfnamefont {R.~D.}\ \bibnamefont {Leonardo}},\ }\bibfield  {title} {\enquote {\bibinfo {title} {Targeted delivery of colloids by swimming bacteria},}\ }\href{\doibase https://doi.org/10.1038/ncomms3588} {\bibfield  {journal} {\bibinfo  {journal} {Nature Communications}\ }\textbf {\bibinfo {volume} {4}},\ \bibinfo {pages} {2588} (\bibinfo {year} {2013})}\BibitemShut {NoStop}%
	\bibitem [{\citenamefont {Liu}\ \emph {et~al.}(2020)\citenamefont {Liu}, \citenamefont {Ou}, \citenamefont {Wang}, \citenamefont {Gao}, \citenamefont {Liu}, \citenamefont {Ye}, \citenamefont {Wilson}, \citenamefont {Hu}, \citenamefont {Peng},\ and\ \citenamefont {Tu}}]{Liu_2020}%
	  \BibitemOpen
	  \bibfield  {author} {\bibinfo {author} {\bibfnamefont {K.}~\bibnamefont {Liu}}, \bibinfo {author} {\bibfnamefont {J.}~\bibnamefont {Ou}}, \bibinfo {author} {\bibfnamefont {S.}~\bibnamefont {Wang}}, \bibinfo {author} {\bibfnamefont {J.}~\bibnamefont {Gao}}, \bibinfo {author} {\bibfnamefont {L.}~\bibnamefont {Liu}}, \bibinfo {author} {\bibfnamefont {Y.}~\bibnamefont {Ye}}, \bibinfo {author} {\bibfnamefont {D.~A.}\ \bibnamefont {Wilson}}, \bibinfo {author} {\bibfnamefont {Y.}~\bibnamefont {Hu}}, \bibinfo {author} {\bibfnamefont {F.}~\bibnamefont {Peng}}, \ and\ \bibinfo {author} {\bibfnamefont {Y.}~\bibnamefont {Tu}},\ }\bibfield  {title} {\enquote {\bibinfo {title} {Magnesium-based micromotors for enhanced active and synergistic hydrogen chemotherapy},}\ }\href{\doibase https://doi.org/10.1016/j.apmt.2020.100694} {\bibfield  {journal} {\bibinfo  {journal} {Applied Materials Today}\ }\textbf {\bibinfo {volume} {20}},\ \bibinfo {pages} {100694} (\bibinfo {year} {2020})}\BibitemShut {NoStop}%
	\bibitem [{\citenamefont {Pan{\'e}}\ \emph {et~al.}(2019)\citenamefont {Pan{\'e}}, \citenamefont {Puigmart{\'\i}-Luis}, \citenamefont {Bergeles}, \citenamefont {Chen}, \citenamefont {Pellicer}, \citenamefont {Sort}, \citenamefont {Po{\v{c}}epcov{\'a}}, \citenamefont {Ferreira},\ and\ \citenamefont {Nelson}}]{pane2019imaging}%
	  \BibitemOpen
	  \bibfield  {author} {\bibinfo {author} {\bibfnamefont {S.}~\bibnamefont {Pan{\'e}}}, \bibinfo {author} {\bibfnamefont {J.}~\bibnamefont {Puigmart{\'\i}-Luis}}, \bibinfo {author} {\bibfnamefont {C.}~\bibnamefont {Bergeles}}, \bibinfo {author} {\bibfnamefont {X.-Z.}\ \bibnamefont {Chen}}, \bibinfo {author} {\bibfnamefont {E.}~\bibnamefont {Pellicer}}, \bibinfo {author} {\bibfnamefont {J.}~\bibnamefont {Sort}}, \bibinfo {author} {\bibfnamefont {V.}~\bibnamefont {Po{\v{c}}epcov{\'a}}}, \bibinfo {author} {\bibfnamefont {A.}~\bibnamefont {Ferreira}}, \ and\ \bibinfo {author} {\bibfnamefont {B.~J.}\ \bibnamefont {Nelson}},\ }\bibfield  {title} {\enquote {\bibinfo {title} {Imaging technologies for biomedical micro-and nanoswimmers},}\ }\href{\doibase https://doi.org/10.1002/admt.201800575} {\bibfield  {journal} {\bibinfo  {journal} {Advanced Materials Technologies}\ }\textbf {\bibinfo {volume} {4}},\ \bibinfo {pages} {1800575} (\bibinfo {year} {2019})}\BibitemShut {NoStop}%
	\bibitem [{\citenamefont {Hosseini}\ \emph {et~al.}(2020)\citenamefont {Hosseini}, \citenamefont {Ahmadi}, \citenamefont {Khoobi}, \citenamefont {Dehghani},\ and\ \citenamefont {Kefayat}}]{hosseini2020high}%
	  \BibitemOpen
	  \bibfield  {author} {\bibinfo {author} {\bibfnamefont {M.}~\bibnamefont {Hosseini}}, \bibinfo {author} {\bibfnamefont {Z.}~\bibnamefont {Ahmadi}}, \bibinfo {author} {\bibfnamefont {M.}~\bibnamefont {Khoobi}}, \bibinfo {author} {\bibfnamefont {S.}~\bibnamefont {Dehghani}}, \ and\ \bibinfo {author} {\bibfnamefont {A.}~\bibnamefont {Kefayat}},\ }\bibfield  {title} {\enquote {\bibinfo {title} {High-performance spirulina--bismuth biohybrids for enhanced computed tomography imaging},}\ }\href{\doibase https://doi.org/10.1021/acssuschemeng.0c04877} {\bibfield  {journal} {\bibinfo  {journal} {ACS sustainable chemistry \& engineering}\ }\textbf {\bibinfo {volume} {8}},\ \bibinfo {pages} {13085} (\bibinfo {year} {2020})}\BibitemShut {NoStop}%
	\bibitem [{\citenamefont {Vyskocil}\ \emph {et~al.}(2020)\citenamefont {Vyskocil}, \citenamefont {Mayorga-Martinez}, \citenamefont {Jablonsk{\'a}}, \citenamefont {Novotny}, \citenamefont {Ruml},\ and\ \citenamefont {Pumera}}]{vyskocil2020cancer}%
	  \BibitemOpen
	  \bibfield  {author} {\bibinfo {author} {\bibfnamefont {J.}~\bibnamefont {Vyskocil}}, \bibinfo {author} {\bibfnamefont {C.~C.}\ \bibnamefont {Mayorga-Martinez}}, \bibinfo {author} {\bibfnamefont {E.}~\bibnamefont {Jablonsk{\'a}}}, \bibinfo {author} {\bibfnamefont {F.}~\bibnamefont {Novotny}}, \bibinfo {author} {\bibfnamefont {T.}~\bibnamefont {Ruml}}, \ and\ \bibinfo {author} {\bibfnamefont {M.}~\bibnamefont {Pumera}},\ }\bibfield  {title} {\enquote {\bibinfo {title} {Cancer cells microsurgery via asymmetric bent surface {A}u/{A}g/{N}i microrobotic scalpels through a transversal rotating magnetic field},}\ }\href{\doibase https://doi.org/10.1021/acsnano.0c01705} {\bibfield  {journal} {\bibinfo  {journal} {ACS nano}\ }\textbf {\bibinfo {volume} {14}},\ \bibinfo {pages} {8247} (\bibinfo {year} {2020})}\BibitemShut {NoStop}%
	\bibitem [{\citenamefont {Yan}\ \emph {et~al.}(2022)\citenamefont {Yan}, \citenamefont {Solovev}, \citenamefont {Huang}, \citenamefont {Cui},\ and\ \citenamefont {Mei}}]{Yan_2022}%
	  \BibitemOpen
	  \bibfield  {author} {\bibinfo {author} {\bibfnamefont {G.}~\bibnamefont {Yan}}, \bibinfo {author} {\bibfnamefont {A.~A.}\ \bibnamefont {Solovev}}, \bibinfo {author} {\bibfnamefont {G.}~\bibnamefont {Huang}}, \bibinfo {author} {\bibfnamefont {J.}~\bibnamefont {Cui}}, \ and\ \bibinfo {author} {\bibfnamefont {Y.}~\bibnamefont {Mei}},\ }\bibfield  {title} {\enquote {\bibinfo {title} {Soft microswimmers: Material capabilities and biomedical applications},}\ }\href{\doibase https://doi.org/10.1016/j.cocis.2022.101609} {\bibfield  {journal} {\bibinfo  {journal} {Current Opinion in Colloid and Interface Science}\ }\textbf {\bibinfo {volume} {61}},\ \bibinfo {pages} {101609} (\bibinfo {year} {2022})}\BibitemShut {NoStop}%
	\bibitem [{\citenamefont {Mahon}\ \emph {et~al.}(2012)\citenamefont {Mahon}, \citenamefont {Salvati}, \citenamefont {{Baldelli Bombelli}}, \citenamefont {Lynch},\ and\ \citenamefont {Dawson}}]{Mahon_2012}%
	  \BibitemOpen
	  \bibfield  {author} {\bibinfo {author} {\bibfnamefont {E.}~\bibnamefont {Mahon}}, \bibinfo {author} {\bibfnamefont {A.}~\bibnamefont {Salvati}}, \bibinfo {author} {\bibfnamefont {F.}~\bibnamefont {{Baldelli Bombelli}}}, \bibinfo {author} {\bibfnamefont {I.}~\bibnamefont {Lynch}}, \ and\ \bibinfo {author} {\bibfnamefont {K.~A.}\ \bibnamefont {Dawson}},\ }\bibfield  {title} {\enquote {\bibinfo {title} {Designing the nanoparticle–biomolecule interface for “targeting and therapeutic delivery”},}\ }\href{\doibase https://doi.org/10.1016/j.jconrel.2012.04.009} {\bibfield  {journal} {\bibinfo  {journal} {Journal of Controlled Release}\ }\textbf {\bibinfo {volume} {161}},\ \bibinfo {pages} {164} (\bibinfo {year} {2012})},\ \bibinfo {note} {drug Delivery Research in Europe}\BibitemShut {NoStop}%
	\bibitem [{\citenamefont {Tottori}\ \emph {et~al.}(2012)\citenamefont {Tottori}, \citenamefont {Zhang}, \citenamefont {Qiu}, \citenamefont {Krawczyk}, \citenamefont {Franco-Obreg{\'o}n},\ and\ \citenamefont {Nelson}}]{Tottori_2012}%
	  \BibitemOpen
	  \bibfield  {author} {\bibinfo {author} {\bibfnamefont {S.}~\bibnamefont {Tottori}}, \bibinfo {author} {\bibfnamefont {L.}~\bibnamefont {Zhang}}, \bibinfo {author} {\bibfnamefont {F.}~\bibnamefont {Qiu}}, \bibinfo {author} {\bibfnamefont {K.~K.}\ \bibnamefont {Krawczyk}}, \bibinfo {author} {\bibfnamefont {A.}~\bibnamefont {Franco-Obreg{\'o}n}}, \ and\ \bibinfo {author} {\bibfnamefont {B.~J.}\ \bibnamefont {Nelson}},\ }\bibfield  {title} {\enquote {\bibinfo {title} {Magnetic helical micromachines: fabrication, controlled swimming, and cargo transport},}\ }\href{\doibase http://dx.doi.org/10.1002/adma.201103818} {\bibfield  {journal} {\bibinfo  {journal} {Advanced materials}\ }\textbf {\bibinfo {volume} {24}},\ \bibinfo {pages} {811} (\bibinfo {year} {2012})}\BibitemShut {NoStop}%
	\bibitem [{\citenamefont {Bunea}\ and\ \citenamefont {Taboryski}(2020)}]{bunea2020recent}%
	  \BibitemOpen
	  \bibfield  {author} {\bibinfo {author} {\bibfnamefont {A.-I.}\ \bibnamefont {Bunea}}\ and\ \bibinfo {author} {\bibfnamefont {R.}~\bibnamefont {Taboryski}},\ }\bibfield  {title} {\enquote {\bibinfo {title} {Recent advances in microswimmers for biomedical applications},}\ }\href{\doibase https://doi.org/10.3390/mi11121048} {\bibfield  {journal} {\bibinfo  {journal} {Micromachines}\ }\textbf {\bibinfo {volume} {11}},\ \bibinfo {pages} {1048} (\bibinfo {year} {2020})}\BibitemShut {NoStop}%
	\bibitem [{\citenamefont {Kurzthaler}\ \emph {et~al.}(2024)\citenamefont {Kurzthaler}, \citenamefont {Zhao}, \citenamefont {Zhou}, \citenamefont {Schwarz-Linek}, \citenamefont {Devailly}, \citenamefont {Arlt}, \citenamefont {Huang}, \citenamefont {Poon}, \citenamefont {Franosch}, \citenamefont {Tailleur} \emph {et~al.}}]{kurzthaler2024characterization}%
	  \BibitemOpen
	  \bibfield  {author} {\bibinfo {author} {\bibfnamefont {C.}~\bibnamefont {Kurzthaler}}, \bibinfo {author} {\bibfnamefont {Y.}~\bibnamefont {Zhao}}, \bibinfo {author} {\bibfnamefont {N.}~\bibnamefont {Zhou}}, \bibinfo {author} {\bibfnamefont {J.}~\bibnamefont {Schwarz-Linek}}, \bibinfo {author} {\bibfnamefont {C.}~\bibnamefont {Devailly}}, \bibinfo {author} {\bibfnamefont {J.}~\bibnamefont {Arlt}}, \bibinfo {author} {\bibfnamefont {J.-D.}\ \bibnamefont {Huang}}, \bibinfo {author} {\bibfnamefont {W.~C.}\ \bibnamefont {Poon}}, \bibinfo {author} {\bibfnamefont {T.}~\bibnamefont {Franosch}}, \bibinfo {author} {\bibfnamefont {J.}~\bibnamefont {Tailleur}},  \emph {et~al.},\ }\bibfield  {title} {\enquote {\bibinfo {title} {Characterization and control of the run-and-tumble dynamics of {E}scherichia {C}oli},}\ }\href{\doibase https://doi.org/10.1103/PhysRevLett.132.038302} {\bibfield  {journal} {\bibinfo  {journal} {Phys. Rev. Lett.}\ }\textbf {\bibinfo {volume} {132}},\ \bibinfo {pages} {038302} (\bibinfo {year} {2024})}\BibitemShut {NoStop}%
	\bibitem [{\citenamefont {Zhiyu}\ \emph {et~al.}(2021)\citenamefont {Zhiyu}, \citenamefont {Jiang},\ and\ \citenamefont {Hou}}]{Zhiyu_2021}%
	  \BibitemOpen
	  \bibfield  {author} {\bibinfo {author} {\bibfnamefont {C.}~\bibnamefont {Zhiyu}}, \bibinfo {author} {\bibfnamefont {H.}~\bibnamefont {Jiang}}, \ and\ \bibinfo {author} {\bibfnamefont {Z.}~\bibnamefont {Hou}},\ }\bibfield  {title} {\enquote {\bibinfo {title} {Designing circle swimmers: Principles and strategies},}\ }\href{\doibase 10.1063/5.0065529} {\bibfield  {journal} {\bibinfo  {journal} {The Journal of Chemical Physics}\ }\textbf {\bibinfo {volume} {155}},\ \bibinfo {pages} {234901} (\bibinfo {year} {2021})}\BibitemShut {NoStop}%
	\bibitem [{\citenamefont {Bechinger}\ \emph {et~al.}(2016)\citenamefont {Bechinger}, \citenamefont {Di~Leonardo}, \citenamefont {L\"owen}, \citenamefont {Reichhardt}, \citenamefont {Volpe},\ and\ \citenamefont {Volpe}}]{Bechinger_2016}%
	  \BibitemOpen
	  \bibfield  {author} {\bibinfo {author} {\bibfnamefont {C.}~\bibnamefont {Bechinger}}, \bibinfo {author} {\bibfnamefont {R.}~\bibnamefont {Di~Leonardo}}, \bibinfo {author} {\bibfnamefont {H.}~\bibnamefont {L\"owen}}, \bibinfo {author} {\bibfnamefont {C.}~\bibnamefont {Reichhardt}}, \bibinfo {author} {\bibfnamefont {G.}~\bibnamefont {Volpe}}, \ and\ \bibinfo {author} {\bibfnamefont {G.}~\bibnamefont {Volpe}},\ }\bibfield  {title} {\enquote {\bibinfo {title} {Active particles in complex and crowded environments},}\ }\href{\doibase 10.1103/RevModPhys.88.045006} {\bibfield  {journal} {\bibinfo  {journal} {Rev. Mod. Phys.}\ }\textbf {\bibinfo {volume} {88}},\ \bibinfo {pages} {045006} (\bibinfo {year} {2016})}\BibitemShut {NoStop}%
	\bibitem [{\citenamefont {Elgeti}\ \emph {et~al.}(2015)\citenamefont {Elgeti}, \citenamefont {Winkler},\ and\ \citenamefont {Gompper}}]{Elgeti_2015}%
	  \BibitemOpen
	  \bibfield  {author} {\bibinfo {author} {\bibfnamefont {J.}~\bibnamefont {Elgeti}}, \bibinfo {author} {\bibfnamefont {R.}~\bibnamefont {Winkler}}, \ and\ \bibinfo {author} {\bibfnamefont {G.}~\bibnamefont {Gompper}},\ }\bibfield  {title} {\enquote {\bibinfo {title} {Physics of microswimmers - single particle motion and collective behavior},}\ }\href{\doibase 10.1088/0034-4885/78/5/056601} {\bibfield  {journal} {\bibinfo  {journal} {Reports on Progress in Physics}\ }\textbf {\bibinfo {volume} {78}},\ \bibinfo {pages} {056601} (\bibinfo {year} {2015})}\BibitemShut {NoStop}%
	\bibitem [{\citenamefont {Liebchen}\ and\ \citenamefont {Levis}(2017)}]{Liebchen_2017}%
	  \BibitemOpen
	  \bibfield  {author} {\bibinfo {author} {\bibfnamefont {B.}~\bibnamefont {Liebchen}}\ and\ \bibinfo {author} {\bibfnamefont {D.}~\bibnamefont {Levis}},\ }\bibfield  {title} {\enquote {\bibinfo {title} {Collective behavior of chiral active matter: Pattern formation and enhanced flocking},}\ }\href{\doibase 10.1103/PhysRevLett.119.058002} {\bibfield  {journal} {\bibinfo  {journal} {Phys. Rev. Lett.}\ }\textbf {\bibinfo {volume} {119}},\ \bibinfo {pages} {058002} (\bibinfo {year} {2017})}\BibitemShut {NoStop}%
	\bibitem [{\citenamefont {K\"ummel}\ \emph {et~al.}(2013)\citenamefont {K\"ummel}, \citenamefont {ten Hagen}, \citenamefont {Wittkowski}, \citenamefont {Buttinoni}, \citenamefont {Eichhorn}, \citenamefont {Volpe}, \citenamefont {L\"owen},\ and\ \citenamefont {Bechinger}}]{Kuemmel_2013}%
	  \BibitemOpen
	  \bibfield  {author} {\bibinfo {author} {\bibfnamefont {F.}~\bibnamefont {K\"ummel}}, \bibinfo {author} {\bibfnamefont {B.}~\bibnamefont {ten Hagen}}, \bibinfo {author} {\bibfnamefont {R.}~\bibnamefont {Wittkowski}}, \bibinfo {author} {\bibfnamefont {I.}~\bibnamefont {Buttinoni}}, \bibinfo {author} {\bibfnamefont {R.}~\bibnamefont {Eichhorn}}, \bibinfo {author} {\bibfnamefont {G.}~\bibnamefont {Volpe}}, \bibinfo {author} {\bibfnamefont {H.}~\bibnamefont {L\"owen}}, \ and\ \bibinfo {author} {\bibfnamefont {C.}~\bibnamefont {Bechinger}},\ }\bibfield  {title} {\enquote {\bibinfo {title} {Circular motion of asymmetric self-propelling particles},}\ }\href{\doibase 10.1103/PhysRevLett.110.198302} {\bibfield  {journal} {\bibinfo  {journal} {Phys. Rev. Lett.}\ }\textbf {\bibinfo {volume} {110}},\ \bibinfo {pages} {198302} (\bibinfo {year} {2013})}\BibitemShut {NoStop}%
	\bibitem [{\citenamefont {L{\"o}wen}(2016)}]{lowen2016chirality}%
	  \BibitemOpen
	  \bibfield  {author} {\bibinfo {author} {\bibfnamefont {H.}~\bibnamefont {L{\"o}wen}},\ }\bibfield  {title} {\enquote {\bibinfo {title} {Chirality in microswimmer motion: From circle swimmers to active turbulence},}\ }\href{\doibase https://doi.org/10.1140/epjst/e2016-60054-6} {\bibfield  {journal} {\bibinfo  {journal} {The European Physical Journal Special Topics}\ }\textbf {\bibinfo {volume} {225}},\ \bibinfo {pages} {2319} (\bibinfo {year} {2016})}\BibitemShut {NoStop}%
	\bibitem [{\citenamefont {ten Hagen}\ \emph {et~al.}(2014)\citenamefont {ten Hagen}, \citenamefont {K{\"u}mmel}, \citenamefont {Wittkowski}, \citenamefont {Takagi}, \citenamefont {L{\"o}wen},\ and\ \citenamefont {Bechinger}}]{BorgetenHagen_2014}%
	  \BibitemOpen
	  \bibfield  {author} {\bibinfo {author} {\bibfnamefont {B.}~\bibnamefont {ten Hagen}}, \bibinfo {author} {\bibfnamefont {F.}~\bibnamefont {K{\"u}mmel}}, \bibinfo {author} {\bibfnamefont {R.}~\bibnamefont {Wittkowski}}, \bibinfo {author} {\bibfnamefont {D.}~\bibnamefont {Takagi}}, \bibinfo {author} {\bibfnamefont {H.}~\bibnamefont {L{\"o}wen}}, \ and\ \bibinfo {author} {\bibfnamefont {C.}~\bibnamefont {Bechinger}},\ }\bibfield  {title} {\enquote {\bibinfo {title} {Gravitaxis of asymmetric self-propelled colloidal particles},}\ }\href{\doibase https://doi.org/10.1038/ncomms5829} {\bibfield  {journal} {\bibinfo  {journal} {Nature Communications}\ }\textbf {\bibinfo {volume} {5}},\ \bibinfo {pages} {4829} (\bibinfo {year} {2014})}\BibitemShut {NoStop}%
	\bibitem [{\citenamefont {Lebert}\ and\ \citenamefont {H{\"a}der}(1999)}]{Lebert_1999}%
	  \BibitemOpen
	  \bibfield  {author} {\bibinfo {author} {\bibfnamefont {M.}~\bibnamefont {Lebert}}\ and\ \bibinfo {author} {\bibfnamefont {D.-P.}\ \bibnamefont {H{\"a}der}},\ }\bibfield  {title} {\enquote {\bibinfo {title} {Negative gravitactic behavior of {E}uglena gracilis can not be described by the mechanism of buoyancy-oriented upward swimming},}\ }\href{\doibase https://doi.org/10.1016/S0273-1177(99)00966-7} {\bibfield  {journal} {\bibinfo  {journal} {Advances in Space Research}\ }\textbf {\bibinfo {volume} {24}},\ \bibinfo {pages} {851} (\bibinfo {year} {1999})},\ \bibinfo {note} {life Sciences: Microgravity Research II}\BibitemShut {NoStop}%
	\bibitem [{\citenamefont {Hemmersbach}\ \emph {et~al.}(2001)\citenamefont {Hemmersbach}, \citenamefont {Bromeis}, \citenamefont {Block}, \citenamefont {Br{\"a}ucker}, \citenamefont {Krause}, \citenamefont {Freiberger}, \citenamefont {Stieber},\ and\ \citenamefont {Wilczek}}]{Hemmersbach_2001}%
	  \BibitemOpen
	  \bibfield  {author} {\bibinfo {author} {\bibfnamefont {R.}~\bibnamefont {Hemmersbach}}, \bibinfo {author} {\bibfnamefont {B.}~\bibnamefont {Bromeis}}, \bibinfo {author} {\bibfnamefont {I.}~\bibnamefont {Block}}, \bibinfo {author} {\bibfnamefont {R.}~\bibnamefont {Br{\"a}ucker}}, \bibinfo {author} {\bibfnamefont {M.}~\bibnamefont {Krause}}, \bibinfo {author} {\bibfnamefont {N.}~\bibnamefont {Freiberger}}, \bibinfo {author} {\bibfnamefont {C.}~\bibnamefont {Stieber}}, \ and\ \bibinfo {author} {\bibfnamefont {M.}~\bibnamefont {Wilczek}},\ }\bibfield  {title} {\enquote {\bibinfo {title} {Paramecium — a model system for studying cellular graviperception},}\ }\href{\doibase https://doi.org/10.1016/S0273-1177(01)00155-7} {\bibfield  {journal} {\bibinfo  {journal} {Advances in Space Research}\ }\textbf {\bibinfo {volume} {27}},\ \bibinfo {pages} {893} (\bibinfo {year} {2001})}\BibitemShut {NoStop}%
	\bibitem [{\citenamefont {Chepizhko}\ and\ \citenamefont {Franosch}(2022)}]{Chepizhko_2022}%
	  \BibitemOpen
	  \bibfield  {author} {\bibinfo {author} {\bibfnamefont {O.}~\bibnamefont {Chepizhko}}\ and\ \bibinfo {author} {\bibfnamefont {T.}~\bibnamefont {Franosch}},\ }\bibfield  {title} {\enquote {\bibinfo {title} {Resonant diffusion of a gravitactic circle swimmer},}\ }\href{\doibase https://doi.org/10.1103/PhysRevLett.129.228003} {\bibfield  {journal} {\bibinfo  {journal} {Phys. Rev. Lett.}\ }\textbf {\bibinfo {volume} {129}},\ \bibinfo {pages} {228003} (\bibinfo {year} {2022})}\BibitemShut {NoStop}%
	\bibitem [{\citenamefont {Kurzthaler}\ and\ \citenamefont {Franosch}(2017)}]{Kurzthaler_2017}%
	  \BibitemOpen
	  \bibfield  {author} {\bibinfo {author} {\bibfnamefont {C.}~\bibnamefont {Kurzthaler}}\ and\ \bibinfo {author} {\bibfnamefont {T.}~\bibnamefont {Franosch}},\ }\bibfield  {title} {\enquote {\bibinfo {title} {Intermediate scattering function of an anisotropic {B}rownian circle swimmer},}\ }\href{\doibase 10.1039/C7SM00873B} {\bibfield  {journal} {\bibinfo  {journal} {Soft Matter}\ }\textbf {\bibinfo {volume} {13}},\ \bibinfo {pages} {6396} (\bibinfo {year} {2017})}\BibitemShut {NoStop}%
	\bibitem [{\citenamefont {Pattanayak}\ \emph {et~al.}(2024)\citenamefont {Pattanayak}, \citenamefont {Shee}, \citenamefont {Chaudhuri},\ and\ \citenamefont {Chaudhuri}}]{pattanayak2024impact}%
	  \BibitemOpen
	  \bibfield  {author} {\bibinfo {author} {\bibfnamefont {A.}~\bibnamefont {Pattanayak}}, \bibinfo {author} {\bibfnamefont {A.}~\bibnamefont {Shee}}, \bibinfo {author} {\bibfnamefont {D.}~\bibnamefont {Chaudhuri}}, \ and\ \bibinfo {author} {\bibfnamefont {A.}~\bibnamefont {Chaudhuri}},\ }\bibfield  {title} {\enquote {\bibinfo {title} {Impact of torque on active brownian particle: exact moments in two and three dimensions},}\ }\href{\doibase 10.1088/1367-2630/ad6a32} {\bibfield  {journal} {\bibinfo  {journal} {New Journal of Physics}\ }\textbf {\bibinfo {volume} {26}},\ \bibinfo {pages} {083024} (\bibinfo {year} {2024})}\BibitemShut {NoStop}%
	\bibitem [{\citenamefont {Kurzthaler}\ \emph {et~al.}(2016)\citenamefont {Kurzthaler}, \citenamefont {Leitmann},\ and\ \citenamefont {Franosch}}]{Kurzthaler_2016}%
	  \BibitemOpen
	  \bibfield  {author} {\bibinfo {author} {\bibfnamefont {C.}~\bibnamefont {Kurzthaler}}, \bibinfo {author} {\bibfnamefont {S.}~\bibnamefont {Leitmann}}, \ and\ \bibinfo {author} {\bibfnamefont {T.}~\bibnamefont {Franosch}},\ }\bibfield  {title} {\enquote {\bibinfo {title} {Intermediate scattering function of an anisotropic active {B}rownian particle},}\ }\href{\doibase 10.1038/srep36702} {\bibfield  {journal} {\bibinfo  {journal} {Scientific Reports}\ }\textbf {\bibinfo {volume} {6}},\ \bibinfo {pages} {36702} (\bibinfo {year} {2016})}\BibitemShut {NoStop}%
	\bibitem [{\citenamefont {Kurzthaler}\ \emph {et~al.}(2018)\citenamefont {Kurzthaler}, \citenamefont {Devailly}, \citenamefont {Arlt}, \citenamefont {Franosch}, \citenamefont {Poon}, \citenamefont {Martinez},\ and\ \citenamefont {Brown}}]{Kurzthaler_2018_janus}%
	  \BibitemOpen
	  \bibfield  {author} {\bibinfo {author} {\bibfnamefont {C.}~\bibnamefont {Kurzthaler}}, \bibinfo {author} {\bibfnamefont {C.}~\bibnamefont {Devailly}}, \bibinfo {author} {\bibfnamefont {J.}~\bibnamefont {Arlt}}, \bibinfo {author} {\bibfnamefont {T.}~\bibnamefont {Franosch}}, \bibinfo {author} {\bibfnamefont {W.~C.~K.}\ \bibnamefont {Poon}}, \bibinfo {author} {\bibfnamefont {V.~A.}\ \bibnamefont {Martinez}}, \ and\ \bibinfo {author} {\bibfnamefont {A.~T.}\ \bibnamefont {Brown}},\ }\bibfield  {title} {\enquote {\bibinfo {title} {Probing the spatiotemporal dynamics of catalytic {J}anus particles with single-particle tracking and {D}ifferential {D}ynamic {M}icroscopy},}\ }\href{\doibase 10.1103/PhysRevLett.121.078001} {\bibfield  {journal} {\bibinfo  {journal} {Phys. Rev. Lett.}\ }\textbf {\bibinfo {volume} {121}},\ \bibinfo {pages} {078001} (\bibinfo {year} {2018})}\BibitemShut {NoStop}%
	\bibitem [{\citenamefont {van Megen}\ and\ \citenamefont {Pusey}(1991)}]{VanMegen_1991}%
	  \BibitemOpen
	  \bibfield  {author} {\bibinfo {author} {\bibfnamefont {W.}~\bibnamefont {van Megen}}\ and\ \bibinfo {author} {\bibfnamefont {P.~N.}\ \bibnamefont {Pusey}},\ }\bibfield  {title} {\enquote {\bibinfo {title} {Dynamic light-scattering study of the glass transition in a colloidal suspension},}\ }\href{\doibase 10.1103/PhysRevA.43.5429} {\bibfield  {journal} {\bibinfo  {journal} {Phys. Rev. A}\ }\textbf {\bibinfo {volume} {43}},\ \bibinfo {pages} {5429} (\bibinfo {year} {1991})}\BibitemShut {NoStop}%
	\bibitem [{\citenamefont {Cerbino}\ and\ \citenamefont {Trappe}(2008)}]{Cerbino_2008}%
	  \BibitemOpen
	  \bibfield  {author} {\bibinfo {author} {\bibfnamefont {R.}~\bibnamefont {Cerbino}}\ and\ \bibinfo {author} {\bibfnamefont {V.}~\bibnamefont {Trappe}},\ }\bibfield  {title} {\enquote {\bibinfo {title} {{D}ifferential {D}ynamic {M}icroscopy: Probing wave vector dependent dynamics with a microscope},}\ }\href{\doibase 10.1103/PhysRevLett.100.188102} {\bibfield  {journal} {\bibinfo  {journal} {Phys. Rev. Lett.}\ }\textbf {\bibinfo {volume} {100}},\ \bibinfo {pages} {188102} (\bibinfo {year} {2008})}\BibitemShut {NoStop}%
	\bibitem [{\citenamefont {Wilson}\ \emph {et~al.}(2011)\citenamefont {Wilson}, \citenamefont {Martinez}, \citenamefont {Schwarz-Linek}, \citenamefont {Tailleur}, \citenamefont {Bryant}, \citenamefont {Pusey},\ and\ \citenamefont {Poon}}]{Wilson_2011}%
	  \BibitemOpen
	  \bibfield  {author} {\bibinfo {author} {\bibfnamefont {L.~G.}\ \bibnamefont {Wilson}}, \bibinfo {author} {\bibfnamefont {V.~A.}\ \bibnamefont {Martinez}}, \bibinfo {author} {\bibfnamefont {J.}~\bibnamefont {Schwarz-Linek}}, \bibinfo {author} {\bibfnamefont {J.}~\bibnamefont {Tailleur}}, \bibinfo {author} {\bibfnamefont {G.}~\bibnamefont {Bryant}}, \bibinfo {author} {\bibfnamefont {P.~N.}\ \bibnamefont {Pusey}}, \ and\ \bibinfo {author} {\bibfnamefont {W.~C.~K.}\ \bibnamefont {Poon}},\ }\bibfield  {title} {\enquote {\bibinfo {title} {{D}ifferential {D}ynamic {M}icroscopy of bacterial motility},}\ }\href{\doibase 10.1103/PhysRevLett.106.018101} {\bibfield  {journal} {\bibinfo  {journal} {Phys. Rev. Lett.}\ }\textbf {\bibinfo {volume} {106}},\ \bibinfo {pages} {018101} (\bibinfo {year} {2011})}\BibitemShut {NoStop}%
	\bibitem [{\citenamefont {Stratonovich}(1963)}]{stratonovich_1967_topics}%
	  \BibitemOpen
	  \bibfield  {author} {\bibinfo {author} {\bibfnamefont {R.~L.}\ \bibnamefont {Stratonovich}},\ }\href{https://books.google.at/books?id=vKkOAAAAQAAJ} {\emph {\bibinfo {title} {Topics in the Theory of Random Noise}}},\ Mathematics and its applications\ (\bibinfo  {publisher} {Gordon and Breach},\ \bibinfo {address} {New York, USA},\ \bibinfo {year} {1963})\BibitemShut {NoStop}%
	\bibitem [{\citenamefont {Gitterman}(2008)}]{gitterman_2008_noisy}%
	  \BibitemOpen
	  \bibfield  {author} {\bibinfo {author} {\bibfnamefont {M.}~\bibnamefont {Gitterman}},\ }\href{https://books.google.at/books?id=UwHGCgAAQBAJ} {\emph {\bibinfo {title} {The Noisy Pendulum}}}\ (\bibinfo  {publisher} {World Scientific Publishing Company},\ \bibinfo {year} {2008})\BibitemShut {NoStop}%
	\bibitem [{\citenamefont {Strogatz}(2018)}]{strogatz2018nonlinear}%
	  \BibitemOpen
	  \bibfield  {author} {\bibinfo {author} {\bibfnamefont {S.~H.}\ \bibnamefont {Strogatz}},\ }\href@noop {} {\emph {\bibinfo {title} {Nonlinear dynamics and chaos: with applications to physics, biology, chemistry, and engineering}}}\ (\bibinfo  {publisher} {CRC press},\ \bibinfo {year} {2018})\BibitemShut {NoStop}%
	\bibitem [{\citenamefont {Risken}(1996)}]{risken1996fokker}%
	  \BibitemOpen
	  \bibfield  {author} {\bibinfo {author} {\bibfnamefont {H.}~\bibnamefont {Risken}},\ }\href{https://doi.org/10.1007/978-3-642-61544-3} {\emph {\bibinfo {title} {Fokker-{P}lanck equation}}}\ (\bibinfo  {publisher} {Springer},\ \bibinfo {year} {1996})\BibitemShut {NoStop}%
	\bibitem [{\citenamefont {Lapolla}\ \emph {et~al.}(2020)\citenamefont {Lapolla}, \citenamefont {Hartich},\ and\ \citenamefont {Godec}}]{Lapolla_2020}%
	  \BibitemOpen
	  \bibfield  {author} {\bibinfo {author} {\bibfnamefont {A.}~\bibnamefont {Lapolla}}, \bibinfo {author} {\bibfnamefont {D.}~\bibnamefont {Hartich}}, \ and\ \bibinfo {author} {\bibfnamefont {A.}~\bibnamefont {Godec}},\ }\bibfield  {title} {\enquote {\bibinfo {title} {Spectral theory of fluctuations in time-average statistical mechanics of reversible and driven systems},}\ }\href{\doibase 10.1103/PhysRevResearch.2.043084} {\bibfield  {journal} {\bibinfo  {journal} {Phys. Rev. Res.}\ }\textbf {\bibinfo {volume} {2}},\ \bibinfo {pages} {043084} (\bibinfo {year} {2020})}\BibitemShut {NoStop}%
	\bibitem [{\citenamefont {Feller}(1991)}]{feller1991introduction}%
	  \BibitemOpen
	  \bibfield  {author} {\bibinfo {author} {\bibfnamefont {W.}~\bibnamefont {Feller}},\ }\href@noop {} {\emph {\bibinfo {title} {An introduction to probability theory and its applications, Volume 2}}},\ Vol.~\bibinfo {volume} {81}\ (\bibinfo  {publisher} {John Wiley \& Sons},\ \bibinfo {year} {1991})\BibitemShut {NoStop}%
	\bibitem [{\citenamefont {Sakurai}\ and\ \citenamefont {Napolitano}(2011)}]{sakurai2011modern}%
	  \BibitemOpen
	  \bibfield  {author} {\bibinfo {author} {\bibfnamefont {J.}~\bibnamefont {Sakurai}}\ and\ \bibinfo {author} {\bibfnamefont {J.}~\bibnamefont {Napolitano}},\ }\href{\doibase https://doi.org/10.1017/9781108499996} {\emph {\bibinfo {title} {Modern Quantum Mechanics}}}\ (\bibinfo  {publisher} {Addison-Wesley},\ \bibinfo {year} {2011})\BibitemShut {NoStop}%
	\bibitem [{\citenamefont {Mayer}\ \emph {et~al.}(2021)\citenamefont {Mayer}, \citenamefont {Sarmiento}, \citenamefont {Escobedo}, \citenamefont {Segovia~Gutiérrez}, \citenamefont {Kurzthaler}, \citenamefont {Egelhaaf},\ and\ \citenamefont {Franosch}}]{Mayer_2021_dimers}%
	  \BibitemOpen
	  \bibfield  {author} {\bibinfo {author} {\bibfnamefont {D.}~\bibnamefont {Mayer}}, \bibinfo {author} {\bibfnamefont {E.}~\bibnamefont {Sarmiento}}, \bibinfo {author} {\bibfnamefont {M.}~\bibnamefont {Escobedo}}, \bibinfo {author} {\bibfnamefont {J.~P.}\ \bibnamefont {Segovia~Gutiérrez}}, \bibinfo {author} {\bibfnamefont {C.}~\bibnamefont {Kurzthaler}}, \bibinfo {author} {\bibfnamefont {S.}~\bibnamefont {Egelhaaf}}, \ and\ \bibinfo {author} {\bibfnamefont {T.}~\bibnamefont {Franosch}},\ }\bibfield  {title} {\enquote {\bibinfo {title} {Two-dimensional {B}rownian motion of anisotropic dimers},}\ }\href{\doibase 10.1103/PhysRevE.104.014605} {\bibfield  {journal} {\bibinfo  {journal} {Physical Review E}\ }\textbf {\bibinfo {volume} {104}},\ \bibinfo {pages} {014605} (\bibinfo {year} {2021})}\BibitemShut {NoStop}%
	\bibitem [{\citenamefont {Narinder}\ \emph {et~al.}(2018)\citenamefont {Narinder}, \citenamefont {Bechinger},\ and\ \citenamefont {Gomez-Solano}}]{narinder2018memory}%
	  \BibitemOpen
	  \bibfield  {author} {\bibinfo {author} {\bibfnamefont {N.}~\bibnamefont {Narinder}}, \bibinfo {author} {\bibfnamefont {C.}~\bibnamefont {Bechinger}}, \ and\ \bibinfo {author} {\bibfnamefont {J.~R.}\ \bibnamefont {Gomez-Solano}},\ }\bibfield  {title} {\enquote {\bibinfo {title} {Memory-induced transition from a persistent random walk to circular motion for achiral microswimmers},}\ }\href{\doibase https://doi.org/10.1103/PhysRevLett.121.078003} {\bibfield  {journal} {\bibinfo  {journal} {Phys. Rev. Lett.}\ }\textbf {\bibinfo {volume} {121}},\ \bibinfo {pages} {078003} (\bibinfo {year} {2018})}\BibitemShut {NoStop}%
	\bibitem [{\citenamefont {Reichhardt}\ and\ \citenamefont {Reichhardt}(2013)}]{reichhardt2013dynamics}%
	  \BibitemOpen
	  \bibfield  {author} {\bibinfo {author} {\bibfnamefont {C.}~\bibnamefont {Reichhardt}}\ and\ \bibinfo {author} {\bibfnamefont {C.~O.}\ \bibnamefont {Reichhardt}},\ }\bibfield  {title} {\enquote {\bibinfo {title} {Dynamics and separation of circularly moving particles in asymmetrically patterned arrays},}\ }\href{\doibase https://doi.org/10.1103/PhysRevE.88.042306} {\bibfield  {journal} {\bibinfo  {journal} {Phys. Rev. E}\ }\textbf {\bibinfo {volume} {88}},\ \bibinfo {pages} {042306} (\bibinfo {year} {2013})}\BibitemShut {NoStop}%
	\bibitem [{\citenamefont {Peng}\ \emph {et~al.}(2016)\citenamefont {Peng}, \citenamefont {Turiv}, \citenamefont {Zhang}, \citenamefont {Guo}, \citenamefont {Shiyanovskii}, \citenamefont {Wei}, \citenamefont {De~Pablo},\ and\ \citenamefont {Lavrentovich}}]{peng2016controlling}%
	  \BibitemOpen
	  \bibfield  {author} {\bibinfo {author} {\bibfnamefont {C.}~\bibnamefont {Peng}}, \bibinfo {author} {\bibfnamefont {T.}~\bibnamefont {Turiv}}, \bibinfo {author} {\bibfnamefont {R.}~\bibnamefont {Zhang}}, \bibinfo {author} {\bibfnamefont {Y.}~\bibnamefont {Guo}}, \bibinfo {author} {\bibfnamefont {S.~V.}\ \bibnamefont {Shiyanovskii}}, \bibinfo {author} {\bibfnamefont {Q.-H.}\ \bibnamefont {Wei}}, \bibinfo {author} {\bibfnamefont {J.}~\bibnamefont {De~Pablo}}, \ and\ \bibinfo {author} {\bibfnamefont {O.~D.}\ \bibnamefont {Lavrentovich}},\ }\bibfield  {title} {\enquote {\bibinfo {title} {Controlling placement of nonspherical (boomerang) colloids in nematic cells with photopatterned director},}\ }\href{\doibase https://doi.org/10.1088/0953-8984/29/1/014005} {\bibfield  {journal} {\bibinfo  {journal} {Journal of Physics: Condensed Matter}\ }\textbf {\bibinfo {volume} {29}},\ \bibinfo {pages} {014005} (\bibinfo {year} {2016})}\BibitemShut {NoStop}%
	\bibitem [{\citenamefont {Giavazzi}\ \emph {et~al.}(2009)\citenamefont {Giavazzi}, \citenamefont {Brogioli}, \citenamefont {Trappe}, \citenamefont {Bellini},\ and\ \citenamefont {Cerbino}}]{Giavazzi_2009}%
	  \BibitemOpen
	  \bibfield  {author} {\bibinfo {author} {\bibfnamefont {F.}~\bibnamefont {Giavazzi}}, \bibinfo {author} {\bibfnamefont {D.}~\bibnamefont {Brogioli}}, \bibinfo {author} {\bibfnamefont {V.}~\bibnamefont {Trappe}}, \bibinfo {author} {\bibfnamefont {T.}~\bibnamefont {Bellini}}, \ and\ \bibinfo {author} {\bibfnamefont {R.}~\bibnamefont {Cerbino}},\ }\bibfield  {title} {\enquote {\bibinfo {title} {Scattering information obtained by optical microscopy: {D}ifferential {D}ynamic {M}icroscopy and beyond},}\ }\href{\doibase 10.1103/PhysRevE.80.031403} {\bibfield  {journal} {\bibinfo  {journal} {Phys. Rev. E}\ }\textbf {\bibinfo {volume} {80}},\ \bibinfo {pages} {031403} (\bibinfo {year} {2009})}\BibitemShut {NoStop}%
	\bibitem [{\citenamefont {Edera}\ \emph {et~al.}(2021)\citenamefont {Edera}, \citenamefont {Brizioli}, \citenamefont {Zanchetta}, \citenamefont {Petekidis}, \citenamefont {Giavazzi},\ and\ \citenamefont {Cerbino}}]{edera2021deformation}%
	  \BibitemOpen
	  \bibfield  {author} {\bibinfo {author} {\bibfnamefont {P.}~\bibnamefont {Edera}}, \bibinfo {author} {\bibfnamefont {M.}~\bibnamefont {Brizioli}}, \bibinfo {author} {\bibfnamefont {G.}~\bibnamefont {Zanchetta}}, \bibinfo {author} {\bibfnamefont {G.}~\bibnamefont {Petekidis}}, \bibinfo {author} {\bibfnamefont {F.}~\bibnamefont {Giavazzi}}, \ and\ \bibinfo {author} {\bibfnamefont {R.}~\bibnamefont {Cerbino}},\ }\bibfield  {title} {\enquote {\bibinfo {title} {Deformation profiles and microscopic dynamics of complex fluids during oscillatory shear experiments},}\ }\href{\doibase https://doi.org/10.1039/D1SM01068A} {\bibfield  {journal} {\bibinfo  {journal} {Soft Matter}\ }\textbf {\bibinfo {volume} {17}},\ \bibinfo {pages} {8553} (\bibinfo {year} {2021})}\BibitemShut {NoStop}%
	\bibitem [{\citenamefont {Giavazzi}\ and\ \citenamefont {Cerbino}(2014)}]{giavazzi2014digital}%
	  \BibitemOpen
	  \bibfield  {author} {\bibinfo {author} {\bibfnamefont {F.}~\bibnamefont {Giavazzi}}\ and\ \bibinfo {author} {\bibfnamefont {R.}~\bibnamefont {Cerbino}},\ }\bibfield  {title} {\enquote {\bibinfo {title} {Digital {F}ourier microscopy for soft matter dynamics},}\ }\href{\doibase https://iopscience.iop.org/article/10.1088/2040-8978/16/8/083001} {\bibfield  {journal} {\bibinfo  {journal} {Journal of Optics}\ }\textbf {\bibinfo {volume} {16}},\ \bibinfo {pages} {083001} (\bibinfo {year} {2014})}\BibitemShut {NoStop}%
	\bibitem [{\citenamefont {Richards}\ \emph {et~al.}(2021)\citenamefont {Richards}, \citenamefont {Martinez},\ and\ \citenamefont {Arlt}}]{richards2021particle}%
	  \BibitemOpen
	  \bibfield  {author} {\bibinfo {author} {\bibfnamefont {J.~A.}\ \bibnamefont {Richards}}, \bibinfo {author} {\bibfnamefont {V.~A.}\ \bibnamefont {Martinez}}, \ and\ \bibinfo {author} {\bibfnamefont {J.}~\bibnamefont {Arlt}},\ }\bibfield  {title} {\enquote {\bibinfo {title} {Particle sizing for flowing colloidal suspensions using flow-{D}ifferential {D}ynamic {M}icroscopy},}\ }\href{\doibase https://doi.org/10.1039/D0SM02255A} {\bibfield  {journal} {\bibinfo  {journal} {Soft Matter}\ }\textbf {\bibinfo {volume} {17}},\ \bibinfo {pages} {3945} (\bibinfo {year} {2021})}\BibitemShut {NoStop}%
	\end{thebibliography}

%

\end{document}